\documentclass[%
 reprint,
 amsmath,amssymb,
 aps,
pra,
]{revtex4-1}

\usepackage{physics}
\usepackage{graphicx}% Include figure files
\usepackage{dcolumn}% Align table columns on decimal point
\usepackage{bm}% bold math
\usepackage[caption=false]{subfig}
\usepackage{lineno}
\nolinenumbers
\usepackage{xcolor}
\begin{document}

%\preprint{APS/123-QED}

\title{Chirp Asymmetry as an analogue of Leptogenesis}% Force line breaks with \\

\author{Matthew Commons}
 \email{mccommons@student.ysu.edu}
\author{Nicole Abend}%
 \email{nkabend@student.ysu.edu}
 \author{Ian M. Jones}%
 \email{imjones@student.ysu.edu}
 \author{Jonathon T. George}%
 \email{jtgeorge01@student.ysu.edu}
  \author{Aaron Weiser}%
 \email{awweiser@student.ysu.edu}
\author{Michael Crescimanno}
\affiliation{Department of Physics, \\Youngstown State University, Youngstown Ohio}
\email{dcphtn@gmail.com}

\date{\today}% It is always \today, today,

\begin{abstract}
The effective conjugation symmetry that arises in the rotating wave frame is the analogue of the discrete symmetry in field theory. Breaking this effective conjugation symmetry leads to asymmetries between up- and down- chirped excitation in quantum optical systems.  We use semiclassical quantum optics theory to describe these processes and
experimentally characterize the asymmetry in the optical response in chirped, two-color saturated absorption spectroscopy (SAS) in an atomic vapor cell. Doing so demonstrates a theoretical and phenomenological correspondence to the simplest model of leptogenesis, the process by which our universe purportedly went from equal amounts of matter and antimater to its present matter excess. The understanding  of the asymmetry as due to a broken discrete symmetry under chirp appears to illuminate the underlying processes responsible for other chirp asymmetries previously noted in the literature.
\end{abstract}

\maketitle

%\tableofcontents
\section{Introduction}
\label{sec:intro}
Due to the frequency specificity of resonant response, continuously tuning a light source through a sample's resonant frequency is a simple experimental method for measuring its fundamental physical parameters. Frequency sweeps slower than internal timescales are used in measurements of line centers, widths, and lineshape asymmetries. Faster frequency sweeps (which we refer as chirped adiabatic optical response (CAP) to differentiate them from non-adiabatic chirp regime as in Fig.\ref{fig:exampleTraces}b) excite other co-operative processes that modulate the observed optical response. 

In general the resulting distortions of the optical response in CAP as a function of the instantaneous detuning are not the same for upchirp and downchirp, where `up' and `down' refer to the chirp direction, i.e. respectively a positive or negative frequency change with time. Thus, the average of the up- and down chirp response with respect to the instantaneous detuning is time even whereas the difference is time odd. We refer to the later as chirp asymmetry.

Some of the physical antecedents and phenomenology of chirp asymmetry in optical response have been long studied. In brief , although generic to many optical processes involving skew spectral densities or dynamic AC stark effects in multiphoton transitions without any underlying spectral density asymmetry, the precise role other fundamental parameters (such as intermediate state lifetimes, for one example) play in the causation or mere amplification of the observed time-odd response has, in our reading of the literature, led to some equivocation. Some of this has been illuminated by analytically solvable models of CAP, but those models typically have excitation envelopes not reproduced in experiment and consequently have a somewhat different phenomenology.  

Our aim here is to clarify and amplify the generality of optical chirp asymmetry by connecting it to the broad phenomenological palette of early universe leptogenesis. After a brief review of chirp asymmetry in the existing literature of CAP (and some related processes), we connect it to basic leptogenesis through a pair of theory models (a population model and optical Bloch equations (OBE)) and an experiment. The models serve to  untangle the essential physical ingredients from simply modulating factors. 

The central thesis is that any observed chirp asymmetry results from the temporal coincidence of (1) a broken (effective) discrete symmetry and (2) a non-stationary process. Further, we find the common phenomenology that the asymmetry varies linearly (at small values) in both the moduli characterizing (1) and (2) and then saturates (not always limiting to 1) at large values.  We include below an experimental demonstration of the phenomena in  chirped, two-color saturated absorption spectroscopy (SAS). We also connect this framework to chiral resolutions (for example, in \cite{viedma}, where parity rather than conjugation symmetry is broken) familiar to chemists, and indicate that it provides an explanatory narrative for the observed asymmetry in chirped, staggered two-color transitions\cite{broers}. 

Leptogenesis refers to a process , usually ascribed to the early universe, by which the universe acquired a non-zero lepton number. The typical assumptions include that the early universe was in a spatially symmetrical state  possessing equal numbers of lepton and anti-leptons early on (thus lepton number 0). Although apparently still in search of its definitive theoretical elucidation in our evolving understanding of particle physics, Sakharov (1967, but as reprinted in English in \cite{sakharov91}) understood that there are three physical antecedents to any process leaving that symmetrical state ({\it i.e.} creating net lepton number). 
These are (1) there must be some process that can make leptons and antileptons in the first place, (2) that the rates for making of leptons and antileptons must differ and that (3) that process must be active during a non-equilibrium phase of the universe's evolution. Since leptons and antileptons are connected by $CP$ it is straightforward to generalize condition (2) to active processes that break a discrete symmetry and for (3) we refer to non-equilibrium not only in the sense of being at a uniform constant temperature, but any situation  which is not time-translationally invariant (not stationary). 

In simple models of leptogenesis (for excellent reviews written for non-specialists read \cite{buchmuller1, buchmuller2}) a heavy Majorana neutrino species falls out of thermal equilibrium as the universe cools, and its subsequent $CP$-violating decays leave a lepton excess. In most of these models the $CP$-violating part of the decay amplitude is usually taken to be proportional to symmetric part. After the production of a lepton excess early in the universe subsequent lower energy processes, usually assumed at the electroweak scale, turn this lepton excess into a hadronic excess and eventually to a universe filled with matter instead of equal parts matter and antimatter. It is a longstanding riddle that in order to arrive at the present matter excess, the $CP$-violating decay amplitude needed in this picture is  much larger than what has been measured in experiment. It is assumed to be sufficiently large in a sector of the theory that, for one reason or another, is not readily accessible to our (comparatively) low energy experiments near the electroweak scale.

Many clever alternative models that purport to create enough $CP$-violation have been proposed and subject to considerable theoretical scrutiny. In \cite{GHOP} it was shown that the first order reflection off standard model phase domain "walls"  is not sufficient to generate the matter asymmetry\cite{kuzmin}. This inadequacy was also explored in  \cite{huet}, as well as a generalization \cite{kajantie} indicating the consequence of the order of the electroweak phase transition later in the early universe. 
In \cite{blanchet}, \cite{mangano} leptogenesis is discussed in a density matrix formulation, a formulation somewhat closer to the optical analogue described in this present article.

There have been various studies of optical effects caused by broken discrete symmetries in a transiently driven out of equilibrium system. For example, Ref.~\cite{avila} studies a population model and includes the effect of 'leakage' to optically inaccessible states, while Ref.~\cite{broers} highlighted the "intuitive" -versus-  "counter-intuitive" two photon absorption chirp ordering effect. That (and some subsequent references, see Ref.\cite{gunaratne}) work  seemed to suggest that the lifetime of the intermediate state is critically important here, as also in \cite{cardman}, figure 6 in the Rydberg atom context. For two-photon transitions,  an analytically solvable theoretical model  sheds light on the 'intuitive' -vs- 'non-intuitive' two photon CAP \cite{carroll1}.  This has been generalized  to a continuum of intermediate states \cite{carroll2}.
There have been additional experimental and theoretical developments for this case \cite{Paspalakis}.
Note also that chirp asymmetry has also been observed in isolated molecular transitions during chirped excitation in which they record resonance width changes but no line center shift at fast chirp rates \cite{mcculloch}.
In that regime they do record 'ringing' (oscillations in the response), indicative of coherence effects not present in the basic leptogenesis model, which from an optical standpoint is essentially a population model. The experiment described below also displays characteristic 'ringing' at large chirp rates, and elsewhere in the literature (for example, see \cite{shwa}) such behavior is generically referred to as the "non-adiabatic regime" of frequency chirp.  

In studying positive and negative chirps, Ref.~\cite{melinger} went beyond populations to a density matrix approach, however at limited chirp speed and not crisply connect their findings to a quantum optical model. In a largely experimental effort, Ref.~\cite{liedenbaum} used a single beam (frequency-) chirp on a molecular transition (doppler free in a molecular beam) to study the asymmetry between up- and down- chirps but did not relate their findings to a broken discrete symmetry of the lineshape. Those experiments appear to have been limited in chirp speed and were unable to demonstrate 
(what we show below as the generic) saturation of the chirp asymmetry with chirp speed. 
Including frequency noise in adiabatic passage\cite{noel} (as we do below) does not break discrete symmetries but does reveal behavior not explained by standard Landau-Zener (LZ) theory, namely that as expected fast passage can be strongly modulated by the noise. 
Chirp asymmetry in multiphoton (STIRAP) Rydberg excitation indicates that peak position and width of a heralded photon shows roughly linear behavior with chirp speed, and that there is a clear difference between positive and negative chirp excitation \cite{zhou} .
That study appears to have precluded further quantification of that difference. In that reference they also graph the optical response peak height as a function of (positive only) chirp, which we also employ as a convenient observable for quantifying chirp asymmetry in the experiment described below. 

The breaking of $C$ symmetry in most of the discussion below is included explicitly as a  constant indicating the intrinsic spectral asymmetry in the optical response. Theory and experiment indicate that this leads directly to the optical response not being chirp symmetric, a fact noted long ago. In Ref.~\cite{xu} they describe a non-linear optical resonator in which the time-reversal asymmetry is traced to an intensity-dependent spectral asymmetry, and is thus an example of the spontaneous breaking of time-reversal symmetry. Experimental perturbations (higher order non-linearities, technical asymmetries) prevent the experiments described in that referernce from having an exactly second order character of a sharp bifurcation devoid of hysteresis, essentially acting actually more as the explicit effective $CP$ violation as described below. 

Spectral asymmetries associated with chirp asymmetry have been observed in single photon transition \cite{djotyan}. Ref.
\cite{simon} studied frequency chirped, single photon response of a quantum dot but did not include a comparison of up- versus down- chirp or associated asymmetry or how it might depend on the chirp speed. They report no asymmetry, though at the end of that article indicate that an additional (effective) conjugation symmetry violating relaxation process should cause there to be an asymmetry between positive and negative chirp. Although they do not chirp but look at transient response in \cite{li1991, li1996}, they do report optical 'ringing' in fast sweep, as reported below.

The chirp asymmetry in two-photon processes is experimentally well characterized \cite{broers, suptitz}. In Refs.~\cite{rangelov2, rangelov1} the authors compare the similar outcomes between the intuitive and non-intuitive excitation sequences against a LZ expectation by solving the OBE. They show that AC Stark shifts play a major role in differentiating between the two excitation sequences. From the point of view of the present note, the AC Stark effects explicitly break the effective conjugation symmetry of the multiphoton excitation, with the breaking being proportional to the intensity of the pulses. 
Alternatively Ref.~\cite{konar} indicates an up- down- chirp asymmetry apparently due to both AC Stark effects (optically non-linear) as well as a spectral asymmetry in the inhomogeneous broadening of the resonance, in this case of a dissolved chromophore. Of ongoing critical interest, there are also experimental observations of time-reversal asymmetric optical response, whose precise microphysical cause (in terms of some fundamental process breaking the discrete symmetry) has not yet been unequivocally determined \cite{gunaratne}. 

In Ref.\cite{torosov1} the connection between spectral asymmetry and the time-reversal symmetry breaking in a multiphoton excitation process was described in significant detail and in analytically solvable theoretical models, the  Demkov–Kunike (DK) \cite{demkov} model and the Carroll and Hioe (CH) (\cite{carroll1, carroll2} and references therein)  model, all of which have both frequency and field amplitude modulation, unlike the model and experiment here which has just chirp. Due to their more complicated intensity envelopes the population differences at low chirp are not linear in the chirp speed and therefore unlike the phenomenology of leptogenesis and of the quantum optical systems studied below. 

Intrinsic SAS lineshape distortions are the explicit (effective) conjugation breaking 'seed' for the chirp asymmetry we report below. Here they are chiefly due to the asymmetric excited hyperfine state density (dominant cause) and asymmetric inhomogenous broadening due to a non-zero Doppler detuning offset (subdominant cause) of the $^{87}$Rb $F=1$ D1 and D2 transitions. 
SAS lineshape distortions caused by light pressure are other interesting, fundamental sources of effective $CP$ violation in SAS, but they are typically a small effect \cite{grimm, oates}. 
Integrated chirp response for a bunch of different scenarios is discussed in Ref. ~\cite{kuznetsova}, but those authors didn't focus on the difference between up and down chirp as we do presently. 
Other references that describe and quantify the various sources of SAS line distortion and asymmetry include Refs.~\cite{borde, park, hirano1, lefloch}, with Ref.~\cite{sharaby} indicating how lineshape asymmetry effects STIRAP. 

Chirp asymmetry may be important in understanding systematics of modulation-based detection of Electromagnetically Induced Transparency (EIT) resonances. 
In EIT it is well known that unequal optical fields and AC stark effects both play important roles in the two photon lineshape asymmetry.
An EIT-related chirp effect appears to have $\sqrt{|\alpha|}$ dependence in the chirp speed $\alpha$ (see Eq.(25) , Ref.\cite{renzoni_lindner_arimondo_1999})  whereas we describe below a linear dependence on chirp speed, clearly depending for example on the sign of the chirp. Although  Ref.~\cite{valente} clearly observe optical ringing in a modulation-based EIT resonance detection scheme, it is unclear how to precisely map their system parameters onto those of the  'non-adiabatic' regime in a frequency chirp.  Ref.~\cite{shwa} also identified optical ringing broadly as a 'non-adiabatic' regime, as also in Ref.~\cite{phillips2} thought to be due to the spectral and temporal overlap of excitation (excited state hyperfine levels) in the atomic vapor. Modulated CPT resonances, important in modern compact secondary frequency standards, have lineshape distortions, many arising from AC stark effects, which through a modulation scheme contribute to clock instability \cite{phillips}. Chirp asymmetry associated with broken $C$ could in principle lead to a CPT resonance demodulation signal at twice the drive frequency. To our knowledge this signal has not been reported, possibly because the modest the effective chirp speeds in the modulation scheme were considerably smaller than the square of the EIT linewidth. 

At first glance one might think that the LZ theory should provide a straightforward understanding of chirp asymmetry. 
For a compact review of LZ (Landau-Zener) transitions in the optical context, see \cite{rubbmark_kash_littman_kleppner_1981}. The simple result is that the probability of system making the (non-adiabatic) jump from one single ground state to a single excited state is $P = \exp(-2\pi |V_{12}|^2/|{\rm d} E/{\rm d} t|/\hbar)$, where $|{\rm d} E/{\rm d} t|\sim |\alpha|$ is the unsigned sweep rate (proportional to what we call the chirp rate)  and $V_{12}$ is the hamiltonian of the matrix element that causes transition itself and thus the splitting at degeneracy. Note that this result is symmetric in the chirp rate $\alpha$, vanishes as an essential singularity (not linear) in the chirp rate at small $\alpha$ (the adiabatic {\it i.e.} equilibrium limit) and goes to $1$ at large $\alpha$. Since that simple LZ model has no broken discrete symmetry we should not be surprised that its phenomenology is so different than that suggested by the leptogenesis paradigm. 

To make connection between LZ and the leptogenesis picture, we could break conjugation symmetry directly via the introduction of additional levels (for partial results in that direction see Refs.~\cite{cardman,Shytov}). However a more definitive correspondence is via the example of a parity broken interband excitation of Ref. \cite{kitamura}. That study is time reversal symmetric, thus $C$-odd terms arise explicitly by the parity of the applied field. 
In the generalized LZ equation they derive chirp speed $\alpha$ is their $E/t$ and  $\epsilon$, a dimensionless parameter explicitly breaking the discrete symmetry, is their $\delta t /t$. With that dictionary, it is clear that that their Eq.(15) and Eq.(16) has the behavior that the chirp asymmetry is proportional to $\epsilon$ (see their graphic 4b, 5b) and, separately, $\alpha$ at small chirp speeds. It is noteworthy that the inclusion of decay processes only weakly affects the standard LZ result, and so is irrelevant for chirp asymmetry, a point made clear by the analysis below of the symmetries in the rotating wave frame\cite{Akulin, Avishai}. 

There are parametric mechanical analog of a LZ process that appears to be somewhat closer to the goals of the study here\cite{shore_gromovyy_yatsenko_romanenko_2009}. %In the appendix we briefly analytically derive the chirp asymmetry in a simple linear-inhomogenous model of swimming. While it again displays roughly the same phenomenology, it is physically distinct from the parametrically excited case pertinent to chirp asymmetry in optics and matter creation in leptogenesis.  
Finally, note that there appears in the literature at least two different dimensionless metrics for quantifying chirp in "rapid adiabatic passage", the ratio of the Rabi frequency to the square root of the chirp speed and the other being chirp speed divided by the square of the decay rate (so, independent of the Rabi frequency). For simplicity and following Ref.~\cite{ernst}, below we quantify our chirp regimes in terms of the later ratio, except where noted. 
 
After describing the discrete symmetries of $T$ and $C$ in a general open quantum system we briefly summarize the most basic model of Leptogenesis, focusing on the role that the Sakharov conditions play in the creation of a matter-antimatter asymmetry. This is followed by a population model for the optical response of a two-level system, in particular showing how its evolution equations map precisely onto those of early universe leptogenesis, the action of the (analogue) discrete symmetries in that model, what optical parameters explicitly break those symmetries and the role of chirp speed. We then describe in detail the four-level semiclassical quantum optics model (and its discrete symmetries) relevant for our experimental realization using chirped adiabatic passage. This is followed by a brief description of our experiment, its experimental results and a comparison of those results to those of early universe leptogenesis, before a concluding discussion that highlights the  relevance of this approach for understanding several previously recorded experimental observations of chirp asymmetry.

\par

\section{Theory: C and T in a framework for open quantum systems}
\label{sec:theory}
Making initially no assumptions about the symmetries of the Hamiltonian underlying the time development of the 
density matrix $\rho$, we expect
\begin{equation}
   \partial_t \rho = -iH(t)\rho + i\rho H^\dagger(t).
   \label{eq:genlRho}
\end{equation}
%To implement symmetry transformations on this equation, consider first the density matrix $\rho_C$ associated with a conjugated evolution: 
%\begin{equation}
%   \partial_t \rho_C = iH^\dagger(t)\rho_C -i\rho_C H(t)
%   \label{eq:genlRhoC}
%\end{equation}
%Conversely, under time reversal the associated density matrix $\rho_T$ satisfies
%\begin{equation}
%   \partial_t \rho_T = iH(-t)\rho_T - i\rho_T H^\dagger(-t)
%   \label{eq:genlRhoT}
%\end{equation}
At this level of discussion there is no reference to space and so no parity transformation, but under the combined symmetries of $C$ and $T$ we find that $TC=CT$ and
\begin{equation}
   \partial_t \rho_{TC} = -iH^\dagger(-t)\rho_{TC} + i\rho_{TC} H(-t) \qquad , 
   \label{eq:genlRhoTC}
\end{equation}
with the notation that $C^\dagger \rho C = \rho_C$ and $T^\dagger \rho T = \rho_T$. 
Now focusing our attention only on density matrices and associated evolution equations for which $TC=1$, that is, 
$\rho_{TC} = \rho$, we 
%learn that 
%\begin{equation}
%   \left(H(t)-H^\dagger(-t)\right) \rho = \rho \left(H^\dagger(t)-H(-t)\right)
%   \label{eq:TCis1}
%\end{equation} 
%We can now combine Eq.\ref{eq:TCis1} with Eq.\ref{eq:genlRho} and Eq.\ref{eq:genlRhoC} to 
recast the evolution equations into $C$-even and $C$-odd combinations of $\rho_C$ and $\rho$, for example
\begin{equation}
   \partial_t(\rho_C+\rho) = -iH(t)^\dagger \rho_C -iH(-t)^\dagger\rho+i \rho_C H(t) + \rho H(-t)
   \label{eq:combined}
\end{equation}
%with a similar equation for the temporal evolution of the C-odd combination $\rho_C-\rho$.
 
%Up to this point we have not placed any requirements on $H$. 
We will be most interested in the case for which the Hamiltonia $H(t)$ and $H(-t)$ are Hermitian where then, 
%This allows us collect terms in Eq.\ref{eq:combined} as 
%\begin{equation}
%   \partial_t(\rho_C\pm\rho) = -\frac{i}{2}\left[H(t)\mp H(-t), \rho_C-\rho\right] - %\frac{i}{2}\left[ H(t)\pm H(-t), \rho_C+\rho\right] \qquad .
%   \label{eq:Ceven_odd}
%\end{equation}
%This indicates that, as expected, $H(t)\pm H(-t)$ are the $CP$ even and odd combinations of the Hamiltonian. Also, since we anticipate any $C$-violation to be small, we expect that the C-odd terms all to be of subleading order compared to the $C$-even terms. Developing Eq.\ref{eq:Ceven_odd} to leading order only,  we find the simpler set, 
\begin{equation}
   \partial_t(\rho_C +\rho) = - \frac{i}{2}\left[ H(t)+ H(-t), \rho_C+\rho\right]
   \label{eq:Ceven}
\end{equation}
\begin{equation}
   \partial_t(\rho_C-\rho) = -\frac{i}{2}\left[H(t)+ H(-t), \rho_C-\rho\right] - \frac{i}{2}\left[ H(t) - H(-t), \rho_C+\rho\right]\qquad .
   \label{eq:Codd}
\end{equation}
in which we have anticipated the C-odd terms to all be of subleading order compared to the $C$-even terms and so have only developed  Eq.~(\ref{eq:genlRho}) to leading non-zero order. This set, Eqs.~(\ref{eq:Ceven}, ~\ref{eq:Codd}), is the recurring motif highlighted through examples below. 
%Finally note that $\rho_C^\dagger$ satisfies the same equation (Eq. \ref{eq:genlRhoC} ) as $\rho_C$, and for the comparing the same off-diagonal matrix elements we actually compare $\rho_C^\dagger$ and $\rho$, but the $\rho_C^\dagger\pm\rho$ satisfies the same Eqs. \ref{eq:Ceven} , \ref{eq:Codd} above. % This is a HUGE run-on sentence. I'm not quite sure how to say it better yet though.....-Matt

\subsection{A simple model of Leptogenesis}
\label{sec:Lepto}
The three key ingredients for establishing a matter excess from an initial symmetrical state are summarized in the "Sakharov conditions". First, the universe must first have a process that creates matter and antimatter (i.e. it must have fundamental L (lepton number), (or B baryon number)) processes), second there must be a conjugation (or charge conjugation+parity) asymmetry in those processes that allows for the net growth of one species (matter) over the other (antimatter). Finally, and most crucially, the system must undergo an out of (thermal) equilibrium excursion, as being in equilibrium forces all the species-specific forward and backward (i.e. time-reversed) reaction currents to have equal magnitudes.  

The simplest scenario for the creation of matter in the early universe is via the creation of a heavy, relatively weakly-coupled right-handed Majorana neutrino that can couple to leptons (electrons, muons, taus) via a charged higgs  satisfying the first criteria of having a fundamental process that violates L conservation (for a modern review of this idea with linkages to the present paper see Ref.\cite{blanchet}). In brief, let $N_N(z)$ be the number density (as measured in a co-moving volume at rest in the cosmological frame) of these heavy states during epoch $z = M/T$ where $M$ is the neutrino mass and $T$ is the temperature of the universe. Generically, we expect $T$ (and thus $z$) to be monotic functions of a cosmological time co-ordinate $t$, but our conclusions do not depend on whether the universe is in a radiation dominated ($T\sim t^{-\frac{1}{2}}$) or matter dominated ($T\sim t^{-\frac{2}{3}}$) phase. 

Being fermions and their own antiparticle, the heavy Majorana neutrinos would achieve an equilibrium density $N_N^{eq}(z)$ at a particular temperature were the universe to expand much slower than their inverse lifetime. Let $D(z)+S(z)$ represent the thermal ensemble time-dilation (w.r.t. cosmological frame time $t$) average of the inclusive (total) decay rate of the neutrinos. The $S(z)$ is the exclusive rate for decay processes that do not result in leptons, whereas the $D(z)$ is that into leptons. 
Let $0<\epsilon<<1$ be a constant branching ratio difference between their decays into leptons instead of antileptons ({\it i.e.} the $CP$-odd part of $D(z)$). Then in rate equations we have, 
\begin{equation}
   \frac{{\rm d}N_N}{{\rm d}z} = -\left(D(z)+S(z)\right)\left(N_N-N_N^{eq}(z)\right)
   \label{eq:leptogenesis1}
\end{equation}
\begin{equation}
   \frac{{\rm d}N_{B-L}}{{\rm d}z} = \epsilon D(z)\left(N_N-N_N^{eq}(z)\right) -W(z)N_{B-L} \qquad .
\label{eq:leptogenesis}
\end{equation}
Here $N_{B-L}$ represents the net lepton number density and included is also a so-called "washout" rate $W(z)$ for the reverse processes. 

Qualitatively, for our purposes, one can understand the behavior of the 
system in terms of two constants, $K$ and $\epsilon$. $K$ is the ratio of the rest frame decay rate of the heavy neutrinos to that of the Hubble parameter $H(z=1)$ value when universe's temperature equals the mass of the heavy neutrino.
Integrating Eqs.~(\ref{eq:leptogenesis1}, \ref{eq:leptogenesis}), one learns that the final lepton asymmetry $N_{B-L}(z=\infty)$ is, for small $1/K$, initially linear in $\epsilon$. The behavior in $K$ is such that for  $1/K \sim 0$ (slow universe expansion) the asymptotic  $N_{B-L}(z=\infty)$ goes to zero as expected since the heavy neutrinos decay away while in thermal equilibrium. $N_{B-L}(z=\infty)$
then increases linearly with $1/K$,  eventually saturating at large $1/K$. In a fast universe expansion each heavy neutrino created in the earlier epoch decays without there being any reverse current from decay products and thus a fixed percentage $\epsilon$ of all the decays lead to the matter excess.   

Before concluding this section we note that the opposite of this generic behavior occurs for processes that tend to drive the system back to its symmetric (discrete symmetry restored) state. In the above example, the change due to washout $W(z)$ in the final asymmetry actually decreases monotonically and saturates (at 0) as $1/K$ increases. This inverse behavior is also seen in more exotic processes, one recent example being the expected reduction of an initial Baryon asymmetry due to the decay of long-lived non-topological solutions (Q balls) that may dominate an early phase of the universe (\cite{white}, Eq.~(10) since $T_{dec}$, being an inverse power of a timescale, is effectively a monotonic function of $\frac{1}{K}$). 

We again refer the interested reader to \cite{buchmuller1, buchmuller2} for a very readable introduction to the leptogenesis. 

\subsection{Optical population model analogue of leptogenesis}
\label{sec:populationsModelling}
To connect the forgoing to optical processes, note that the expansion rate of the universe there is both the cause and the measure of how far the process is out of equilibrium. Technically temporal equilibrium is the 'time' version of the spatial symmetries of homogeneity and isotropy, that is, it is a state that is both time translation and time reversal invariant. 
In the optical context chirp speed is an adequate metric of a processes' 'distance' from equilibrium. 

The second ingredient is the (explicit) $CP$ violating processes in the early universe. In the optical context, in the rotating frame $C$-conjugation is the transformation $\Delta \rightarrow -\Delta$ where $\Delta$ is the detuning from the relevant transition. 
Thus, a $CP$ violating process 
in the optical context has a spectral asymmetry. 
In practice the spectral asymmetry can be explicit (a fixed lineshape asymmetry), spontaneous (associated with an endogenous field whose average value is ordinarily zero) or field-driven, for example as in an AC Stark lineshape asymmetry which plays a starring role in the pulsed two-color chirp asymmetry of, for example,  Ref.~\cite{broers}.

The most straightforward and explicit connection 
of an optical model with 
with Eqs.~(\ref{eq:leptogenesis1}, \ref{eq:leptogenesis}) is then via a population model for a non-interacting collection of two level systems. The upper-most state of each $\ket{0}$, is an excited electronic state having a non-zero dipole matrix elements with the ground state $\ket{1}$. 
Let $N_0, \, N_1$ represent the populations of each level, and note that, being closed, let $N_0+N_1 = 1$. Thus an optical field of Rabi frequency $B$ causes transitions between level `1' and level `0' (with level `0' undergoing stimulated transitions to level `1' and spontaneous decay at a rate $\gamma$) so that that the ensuing rate equation for this system reads
\begin{equation}
   \frac{{\rm d}N_1}{{\rm d}t} = \gamma N_0 - {\cal B}(\Delta)(N_1-N_0)
\label{eq:populations}
\end{equation}
Let $\gamma_2>\gamma/2$ be the decay rate of the coherence between the states. For a laser field with constant amplitude $B$ and a slowly changing detuning $\Delta = \alpha t$, one finds that approximately ${\cal B}(\Delta) = 2\gamma_2\rho(\Delta)\frac{|B|^2\gamma_2}{\Delta^2 + \gamma_2}$, where one should regard $\rho(\Delta)$ a modulation of the excitation rate ${\cal B}(\Delta)$ by the local density of states of the excited state manifold `0'.  Note that if `0' is a single quantum state then $\rho(\Delta) = 1$. 
%For CAP in simple systems there is some question as to what is most natural dimensionless chirp parameter. 
%Although the ratio of $|\alpha|/|B|^2$ is sometimes used to parameterize the process, we use the experimentally simpler $\alpha/\gamma^2$ first introduced by Ernst \cite{ernst}, thus in the evaluation of the models (namely Eq.~(\ref{eq:populations}) and Eq.~(\ref{eq:rho00}-\ref{eq:rho31}) below) we scale parameters so as to take $\gamma=1$. 

As described in Ref.~\cite{torosov2}, $C$ is wavefunction conjugation, which in the rotating wave frame is an inversion of (all) detunings about 0, that is , $\Delta \rightarrow -\Delta$. Herethis relates the system experiencing an upchirp $\alpha > 0$ and $\rho = \rho(\Delta)$ to that under a downchirp $\alpha < 0$ and $\rho = \rho(-\Delta)$. Parity symmetry does not play much of a role in population model, since it only flips the sign of the optical field amplitude $B$ to $-B$ and changes no other parameters in Eq.~(\ref{eq:populations}).

We now double the system in Eq.~(\ref{eq:populations}), referring to the downchirp version of the system by `$\bar\ $', as it were,  the "anti-" system. Note that the equations Eqs.~(\ref{eq:leptogenesis1},\ref{eq:leptogenesis}) are in terms of one $CP$-even quantity ($N_N$) and one $CP$-odd quantity ($N_{B-L}$), so we algebraically re-organize our doubled system into a $CP$-even and $CP$-odd equation as 
\begin{equation}
  \frac{{\rm d}S}{{\rm d}t} =  -({\cal B}+{\bar{\cal B}}+\gamma)S +2\gamma
\label{eq:populations2b}
\end{equation}
\begin{equation}
 \frac{{\rm d}R}{{\rm d}t} = -({\cal B}-{\bar{\cal B}})S - 
({\cal B}+{\bar{\cal B}}+\gamma)R
\label{eq:populations2} 
\end{equation}
where $S = N_1+{\bar N}_1-N_0-{\bar N}_0$ and $R =  N_1-{\bar N}_1-N_0+{\bar N}_0$ are the $CP$-even and -odd (respectively) population differences and where in equation Eq.~(\ref{eq:populations2b}) we dropped a product of two $CP$ odd terms since we expect the $CP$ odd terms each to be small and thus keep only the leading order terms. It is now relatively straightforward to cast  Eqs.~(\ref{eq:populations2b}, \ref{eq:populations2}) into the form of Eqs.~(\ref{eq:leptogenesis1}, \ref{eq:leptogenesis}) by a taking $f(t) N_{B-L} = R$ and  $N_N-N_N^{eq}(t) = 2-S$ where the function $\epsilon f(t) = \frac{{\cal B}-{\bar{\cal B}}}{\gamma+{\cal B}+{\bar{\cal B}}}$ 
With these substitutions we recover Eq.~(\ref{eq:leptogenesis1}) and Eq.~(\ref{eq:leptogenesis}) with the identification 
\begin{equation}
 \frac {{\rm d}N^{eq}_N}{{\rm d}t} = -2 ({\cal B}+{\bar{\cal B}})
\label{eq:matching1} 
\end{equation}

Some intuition into the commonalities and differences between the optical process modelled via populations and leptogenesis can be gleaned 
by comparing asymptotic values. In this simple population model the source of the $C$-violation is the non-symmetric (under inversion $\Delta \rightarrow -\Delta $) $\rho(\Delta)$. Assuming its non-symmetry is small, we expand, $\rho(\Delta) = 1+\epsilon \Delta + \ldots$, so that ${\cal B}(t) = (1+\epsilon \alpha t)\frac{2|B|^2\gamma_2}{\alpha^2 t^2 + \gamma_2^2}$, leading to, from Eq.~(\ref{eq:matching1}), 
\begin{equation}
 N^{eq}_N(t) = \frac{4|B|^2}{\alpha\gamma_2}\left[
 \frac{\pi}{2} - \arctan \left(\frac{\alpha t}{\gamma_2} \right)\right]
\label{eq:matching2} 
\end{equation}
which is qualitatively similar to the $N^{eq}_N(z)$ in the minimal models of leptogenesis. Checking also the analogue of $D(z)$ and $W(z)$ that arise in this population model, we find further similarity with the optical population model. Comparing Eq.~(\ref{eq:leptogenesis1}) and Eq.~(\ref{eq:populations2}) with the aforementioned variable changes indicates that the $D(z)/z = \gamma + {\cal B}+{\bar{\cal B}}$ whereas $W(z)/z = D(z)/z + \frac{{\rm d}\ln(f)}{{\rm d}t}$. Since $D(z)$ in leptogenesis is related to the thermal (relativistic wrt the cosmic rest frame) averaged decay constant $t$ (i.e. $z$) in the the function $D(z)/z$ approaches a constant at large times, as does our analogue in the population model. Also note that for our $f(t)$ in a linear chirp we expect that $\frac{{\rm d}\ln(f)}{{\rm d}t} \sim -1/t$ at large times, indicating that the washout in this model does not vanish at large $z$. 
In models of leptogenesis the washout function $W(z)/z \rightarrow 0$ as $z\rightarrow \infty$, thus ``freezing in" the matter-antimatter asymmetry for all times. Simply as a consequence of the intrinsic relaxation processes in this optical analogue, our model corresponds to leptogenesis in which the created lepton themselves (and thus the asymmetry $N_{B-L}$) ultimately decay away.

\subsection{Semiclassical Quantum Optics model of the experiment}
\label{sec:QO}

The appearance of this universal form Eqs.~(\ref{eq:leptogenesis1}, \ref{eq:leptogenesis}) emerging from a generalized 2-level population model for an optical system is no surprise, but follows as the most general form with these symmetries for a system of only two degrees of freedom and $CP$ violation. Furthermore, note that the change of ${\cal B}(\Delta)$ under $\Delta \rightarrow -\Delta$, the conjugation symmetry, is the term that drives the asymmetry. Systems symmetric in $\Delta$ have no chirp asymmetry. 
In an effort to connect with experiment we now move beyond the population model, explicitly including the evolution of optical coherences. 

Before understanding the more complicated multilevel model appropriate for the experiment, consider its truncation to just two levels (labeled here as in the population model). 
A two level quantum optics model is contained in the truncation of the system below to just Eq.~(\ref{eq:rho00}) and Eq.~(\ref{eq:rho01}). There note that the discrete transformations read $C$:$\delta \rightarrow -\delta$, $\rho\rightarrow \rho^\dagger$ and $P$: $B\rightarrow -B$ and thus  Eq.~(\ref{eq:rho00}) and Eq.~(\ref{eq:rho01}) are manifestly unchanged under $CP$, for any values of the decay parameters $\gamma$, $\gamma_2$. Now, including a linear chirp, $\delta = \alpha t$, then solving just these equations in time for the $T$-odd pair of up- $(\alpha = |\alpha|)$ and down-chirps $(\alpha = -|\alpha|)$ and comparing the solutions, we again see that due to $CP$ invariance the $\rho(t)$ will depend on $|\alpha|$ and $t$ but not on the sign of $\alpha$, indicating there can be no chirp asymmetry for any $\gamma, \gamma_2$ for any chirp speed $|\alpha|$ even including coherences. It is also not hard to show also that beyond the rotating wave approximation (RWA) in the two-level model from Eq.~(\ref{eq:rho00}) and Eq.~(\ref{eq:rho01}) the optical response is still chirp symmetric. On the other hand, for a dipole transition in an isolated three level system higher order AC Stark contributions will in general break this effective conjugation symmetry and lead to chirp asymmetry. In optical systems with a more complex level structure or including transverse optical modes and multiple laser fields undergoing frequency excursions there are additional sources of (explicit, effective) $CP$ breaking. In this sense chirp asymmetry is expected to be generic in realistic multilevel non-linear optical processes.

\begin{figure}
\includegraphics[width=2.0in]{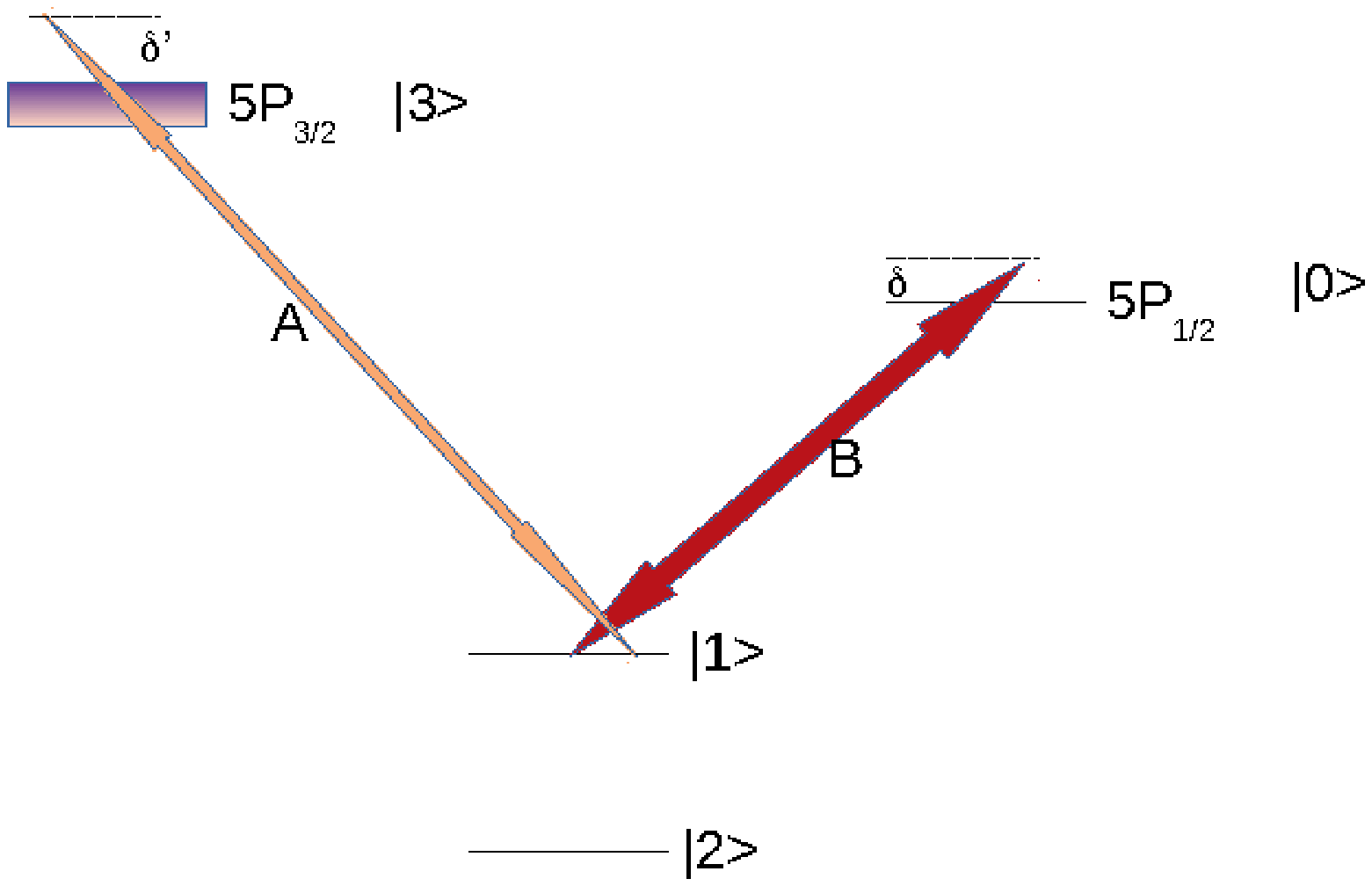}
\includegraphics[width=2.8in]{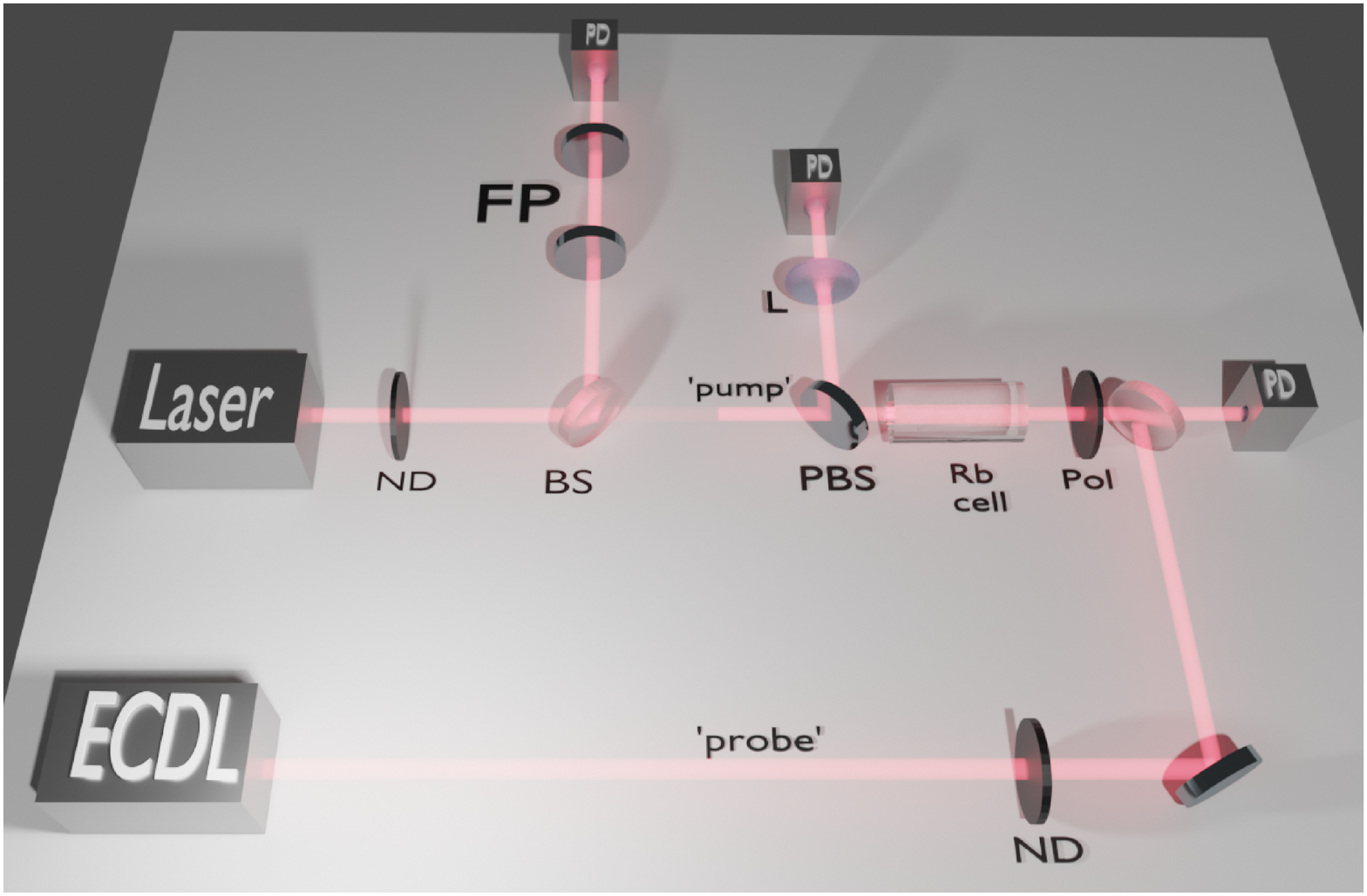}
\caption{(a) The four level, two color saturated absorption model. $A$ is a weak (counter-propagating) 'probe' field of fixed intensity and detuning, whereas $B$ is a bright 'pump' field whose detuning $\delta$ is swept. The states $\ket{1}$ and $\ket{2}$ are the two widely separated 'ground state' hyperfine manifolds of $^{87}$Rb : the hyperfine splitting in the D1 transitions (to $\ket{0}$) are also large and we've not drawn but a single participating state. For the D2 transition the excited state hyperfine levels are significantly closer to one another and, via Doppler detunings, contribute as a manifold $\ket{3}$. (b) Experimental schematic. ND neutral density filter, PBS polarizing beam splitter, ECDL extended cavity diode laser at 780nm, Pol linear polarizer.} 
\label{QOmodel}
\end{figure}

A straightforward experimental realization of this optical analogue of $CP$ violation and leptogenesis in a optical process is via two color saturation absorption spectroscopy. We describe our experimental setup and results later, realized in a four-level quantum optical system we now describe theoretically. The necessity of using the four level system (relevant Rubidium-87 atom levels shown in Fig.\ref{QOmodel}), was thrust upon us by detector gain-bandwidth limitations as described in the experimental section following. Although much of this theory discussion is rudimentary, we include it here for completeness, as it allows us to very concretely describe the analogue $CP$ symmetry of the model and the physics that breaks it. 
 
For Fig.\ref{QOmodel} in practice $A$ (probe) and $B$ (pump) are of very different magnitudes and we numerically solve the semi-classical optical Bloch equations below. Ignoring higher order effects associated with non-resonant light couplings (such as AC Stark effects) and thus working in the rotating wave approximation, we have
\begin{equation}
 \partial_t\rho_{00}= -\gamma\rho_{00} - iB(\rho_{10} - \rho_{01})
 \label{eq:rho00} 
\end{equation}
\begin{equation}
\partial_t\rho_{01}= -(\gamma_2 + i\delta)\rho_{01} - iB(\rho_{11} - \rho_{00})
 \label{eq:rho01} 
\end{equation}
\begin{equation}
\partial_t\rho_{11} = \frac{\gamma}{2}\rho_{00} - \Gamma(\rho_{11} - \rho_{22}) - iB(\rho_{01} - \rho_{10})- iA(\rho_{31} - \rho_{13})
 \label{eq:rho11} 
\end{equation}
%\begin{equation}
%\partial_t\rho_{22} = \frac{\gamma}{2}\rho_{00} - \Gamma(\rho_{22} - %\rho_{11})
%\label{eq:rho22}
%\end{equation}
\begin{equation}
 \partial_t\rho_{33}= -\gamma\rho_{33} - iA(\rho_{13} - \rho_{31})
 \label{eq:rho33} 
\end{equation}
\begin{equation}
\partial_t\rho_{31} = -(\gamma_2 + i\delta')\rho_{31} + iA(\rho_{33} - \rho_{11})
 \label{eq:rho31} 
\end{equation}
%\begin{equation}
%\partial_t\rho_{02} = -\gamma_2\rho_{02} - i\Omega\rho_{12}
%\end{equation}
%\begin{equation}
%\partial_t\rho_{12} = -(\Gamma_2 + i\Delta)\rho_{12} - i\Omega\rho_{02}
%\label{eq:rho12}
%\end{equation}
along with $\rho^\dagger = \rho$, also  $Tr(\rho)=1$ and $\rho_{32} = \rho_{02}=\rho_{12}=0$. 
For our application, the relevant discrete symmetries of the systems of  
equations Eqs.~(\ref{eq:rho00} - \ref{eq:rho31}) are $P$ (parity) :  $A,B \rightarrow -A, -B$ and  $C$ (charge conjugation): $\delta, \delta' \rightarrow -\delta, -\delta'$ along with $\rho \rightarrow \rho^\dagger$
and finally $T$ (time reversal) : $t\rightarrow -t$. 

\begin{figure}
\includegraphics[width=\linewidth]{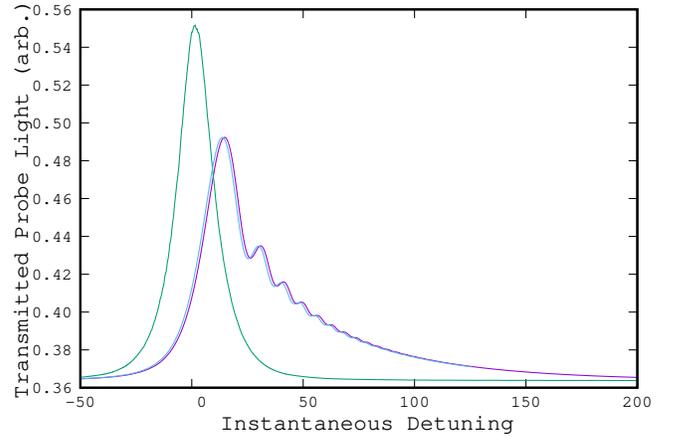}
\caption{A $C$-symmetric theory plot of the transmitted D2 (probe) light ($\delta'=0$) through the optically thin cell from integration of the Bloch equations Eqs.~(\ref{eq:rho01} - \ref{eq:rho31}) during a moderately fast frequency chirp on the D1 (pump). The gold/green trace is actually two traces (up- and down- chirp) without laser frequency noise, at a very slow chirp speed. The blue/purple trace and is also both up- and down- chirp traces (slightly horizontally displaced for clarity) but at a chirp speed of 20 (=$|\alpha|/\gamma^2$). All traces shown include Doppler averaging (Doppler width of 50$\gamma$) and laser FM noise averaging using an 8$\gamma$-wide gaussian distribution} 
\label{fig:theoryTrace}
\end{figure}

\begin{figure}
\includegraphics[width=\linewidth]{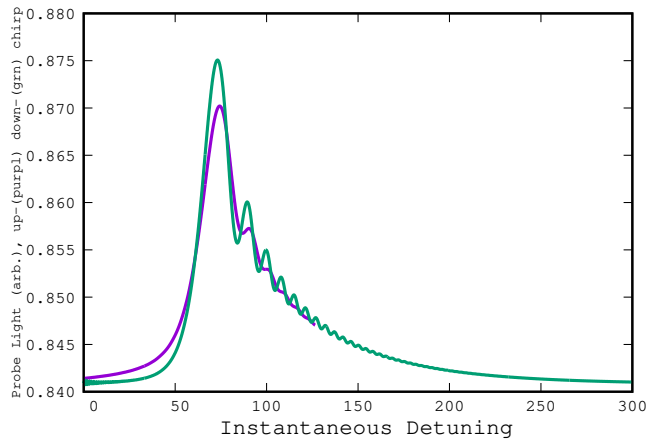}
\caption{A theory plot of the transmitted D2 (probe) light through the optically thin cell from integration of the Bloch equations Eqs.~(\ref{eq:rho01} - \ref{eq:rho31}) using the same  Doppler and laser width and same pump chirp speed as in Fig.\ref{fig:theoryTrace}, but with $\delta'=60$. The non-zero $\delta'$ probe detuning breaks the pump spectral reflection symmetry (the effective $C$-symmetry), leading to chirp asymmetry. Purple is up-chirp and green is down-chirp. Quantifying the chirp asymmetry in this model as a function of the chirp speed is Fig.\ref{fig:asymTheory}a} 
\label{fig:upDownCompareTheory}
\end{figure}

To emphasize the symmetry properties, specialize now to the case where there is but one optically accessible state in the $\ket{3}$ manifold. 
Then, keeping $\delta'$ fixed, under a linear sweep $\delta = \alpha t$ we register the optical response as a function of the detuning $\delta$ and not the time, we thus see that the equation set itself is invariant under the combined action of $CP$ only when $\delta' = 0$ (Note Fig.\ref{fig:theoryTrace}). $CP$ is thus explicitly broken when $\delta' \ne 0$ (note Fig.  \ref{fig:upDownCompareTheory}). 

For more realistic contact with the experiment using a warm vapor cell, our numerical evaluation of this model includes a Boltzmann average over the Doppler shifts associated with different velocities of atoms. Additionally we have included random laser frequency excursions to mimic the finite laser linewidth of the pump beam laser (while the probe laser in our experiments is from an ECDL and so we assume the probe's frequency noise is comparatively negligible, see below for more on the experimental details). 
% set ylabel "Transmitted Probe Light (arb.)" font "computer modern, 32pt" offset -0.5
% set xlabel "Instantaneous Detuning" font "computer modern, 32pt" offset 1
%  set lmargin -8
%  set bmargin 5
% unset key
% set format x "%3.0f"
% set format y "%1.2f"
% set xtics font "computer modern, 24pt" scale 2
% set ytics font "computer modern, 24pt" scale 2
% plot [-50:200] "NoisePos_20" using ($1+72):(100*(0.01-$2)) w l,  "NoisePos_1" using ($1+72):(100*(0.01-$2)) w l, "NoiseNeg_20" using (-145-$1+72):(100*(0.01-$2)) w l
% set terminal postscript eps size 6.5,4.62 enhanced color font "Times Roman,20" linewidth 3
%  set output 'fig2.eps'
% replot

The pump beam spatially enveloped the counter-propagating probe beam, so as the pump reaches resonance, the atoms that interrupt the probe beam may have already been pumped into D1 excited states OR to the lower ground state $\ket{2}$. This produces a transparency spike on the probe (depicted in Fig.\ref{fig:theoryTrace} from theory and Fig.\ref{fig:exampleTraces} from experiment). For that Fig.\ref{fig:theoryTrace}, since the probe detuning is fixed at zero ($\delta'=0$) and no other $C$ violating terms yet added to the system of Eqs.~(\ref{eq:rho01} - \ref{eq:rho31}), then regardless of the sweep speed the optical response (recorded as a function of the detuning) is identical for both up- and down- chirps, as expected. 

Many physically-motivated additions to the set  Eqs.~(\ref{eq:rho01} - \ref{eq:rho31}) either explicitly or inductively break $C$ symmetry. As suggested in the earlier section under the population model, an asymmetric density of states (DOS) intrinsic to one of the participating manifolds can explicitly break $C$. Induced $C$ breaking is caused by the addition of an external field and thus tunable and vanishing in the zero-field limit. An example of induced $C$-breaking are AC Stark effects important for understanding the difference between the so-called "intuitive" and "non-intuitive" two color absorption in CAP. In our experiment the asymmetric DOS of one of the levels (occasioned by the excited state $\ket{3}$ hyperfine manifold)  and Doppler detuning of the probe relative to the zero velocity class are both sources of explicit effective $C$ violation.

Evaluation of the above theory with the addition of the explicit $C$ violation afforded by including either an asymmetric DOS in the manifold $\ket{3}$ or detuning the probe field from the center of the transition between $\ket{1}$ and $\ket{3}$ ({\it i.e.} $\delta'\ne 0$) leads to an asymmetric lineshape at slow chirp that is, again, independent of the sweep direction (i.e. up- or down- chirp). Now, however, increasing the magnitude of the chirp rate, a striking difference appears between the up- and down- chirp traces (theory Fig.\ref{fig:upDownCompareTheory}, experiment Fig.\ref{fig:exampleTraces}). This is phenomenology strongly reminiscent of the Sakharov conditions in that a simultaneous $C$ violation and an out-of-equilibrium condition is necessary to create non-zero lepton number. Optically, once the effective $CP$-symmetry is broken there are multiple differences between up- and down- chirp response, and it is not appriori obvious how best to quantify the difference. For simplicity in what follows we measure for chirp asymmetry the difference in the up- and down- response maxima divided by the sum. 

% set ylabel "Probe Light (arb.), up-(purpl) down-(grn) chirp" font "computer modern, 20pt" offset -0.5
% set xlabel "Instantaneous Detuning" font "computer modern, 32pt" offset 1
%  set lmargin -8
%  set bmargin 5
% r1 = .06 
% s1 = -130
% unset key
% set format x "%3.0f"
% set format y "%0.3f"
% set xtics font "computer modern, 24pt" scale 2
% set ytics font "computer modern, 24pt" scale 2
% plot [0:300] "noiseUP_at_60" using (-$1-72):(1-$2) w l lw 3, "noiseDOWN_at_60" using ($1+190):(1-$2) w l lw 3
% set terminal postscript eps size 6.5,4.62 enhanced color font "Times Roman,20" linewidth 3
%  set output 'fig3.eps'
% replot

\section{Experiment}
\label{sec:exp}
A convenient experimental protocol for chirp asymmetry can be realized through two-color saturated absorption spectroscopy of Rubidium-87 which approximately has a level structure of Fig.\ref{QOmodel}a) and in which the pump and probe fields are tuned to the D1 and D2 transitions respectively. The probe and pump beams thus originate in different lasers and only the pump is frequency swept while the probe is maintained at a weakly locked ($|\delta'|$<30 MHz) fixed frequency inside the doppler envelope of the hyperfine manifold of the vapor cell. The pump and probe overlap and counter propagate inside the natural abundance Rubidium vacuum vapor cell held at 40$^o$C \cite{SASVapor}. The pump beam had a power of less than a milliwatt and a diameter of about 0.8mm whereas the probe beam is typically some 10-30 times weaker but with about 3/4 the diameter of the pump beam. This means that for the large probe field detection bandwidth (below) we used, the probe light level was only (roughly) five times the shot noise limit. 

\begin{figure}
\includegraphics[width=2.6in]{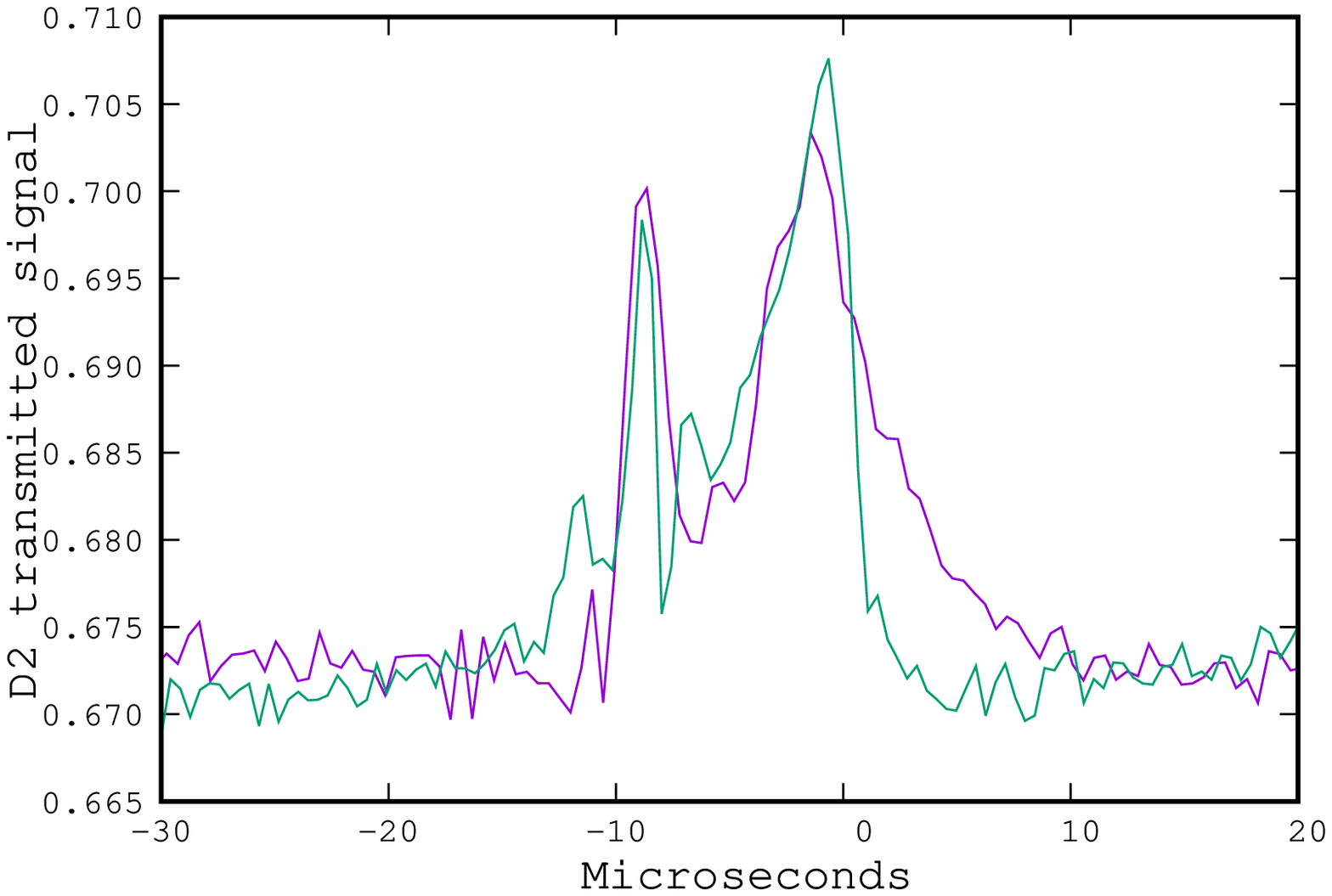}
\includegraphics[width=2.6in]{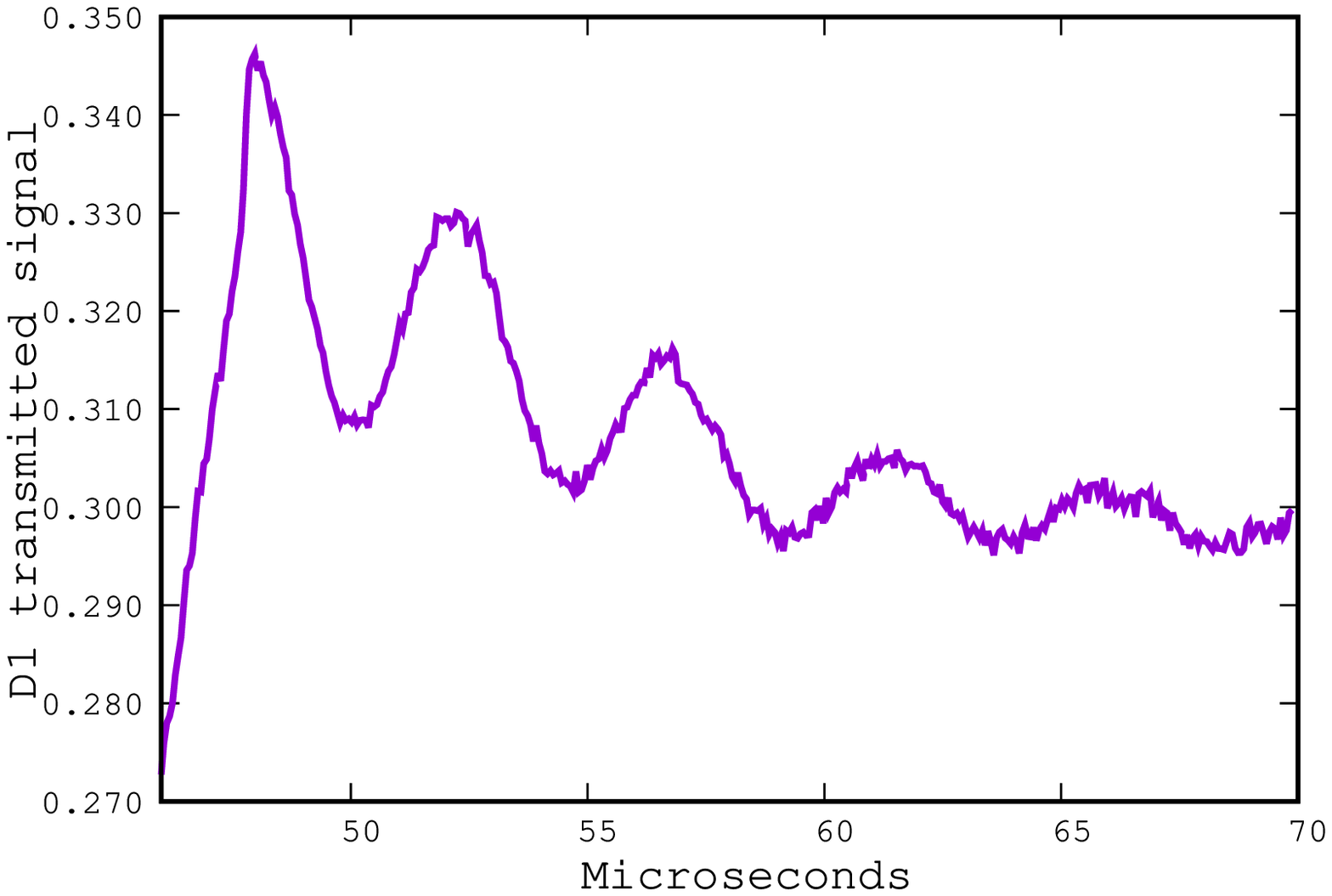}
\caption{(a) Typical D1 (795nm) pump chirp causing transmission changes in the D2 (780nm) counterpropagating probe field, overlaid (purple) up- and (green) down-chirp signals to show the asymmetry. These data were at a sweep of nearly 100 MHz/$\mu$S. (b) At higher chirp speeds the probe transmission has a strong 'ringing' character. For those data the probe was also D1 from a separate ECDL tuned to $^{85}$Rb F=3, the individual excited state hyperfine states are not resolvable and the D1 pump was downchirped at 280 MHz/$\mu$S.} 
\label{fig:exampleTraces}
\end{figure}

In practice, because of the high chirp speeds we require it was impractical to use a ECDL as pump. Instead the swept pump was created by current sweeps in a temperature-tuned mode-hop free, free-running laser diode. Although spectrally wider than ECDLs, it is well known that using free running laser diodes leads to adequate SAS feature definition at slow sweep speeds, and here report that free-running laser diode emmision was spectrally narrow enough to provide the D1 pump field at these higher chirp speeds \cite{wizemann, bucci}. In light of the reported GHz bandwidths for the electronic relaxation timescales in these diodes this is not surprising.  We have used both D1 and D2 ECDLs as weak probe fields in this experiment, but all the data (except Fig.\ref{fig:exampleTraces}b) discussed below used the D2 ECDL for the probe field. 

To facilitate the ready interpretation of the atom's optical response during a fast chirp it is convenient to use amplified photodetectors of large bandwidth. Of note, standard laboratory amplified photodiodes typically have few MHz bandwidths which at these chirp speeds would have introduced additional electronic artifacts making them unsuitable for this experiment. We chose instead to use a fast amplified photodetector  FPD610-FS-NIR  (MenloSystems \textsuperscript{\texttrademark}, with gain=2x$10^6$  and a DC-600 MHz (3dB) bandwidth) to monitor changes in the weak probe beam's transmission.  Because the FPD610-FS-NIR has such a small active area we used a 2cm focal length lens to condense the transmitted probe beam into the detector. With this large detection bandwidth we anticipated and were able to independently verify that any measured signal changes at high chirp speeds were a consequence of the atomic and not electrical processes. Although in principle one could measure a chirp asymmetry interrogating a doppler broadened excited state hyperfine manifold alone (without SAS), the associated detection bandwidth requirements and chirp speeds could not be accommodated in our laboratory even with the FPD610-FS-NIR. The SAS resonances, having few MHz bandwidths, worked well for the laser drivers and the fast detection electronics we used.  Finally we note that these experiments were carried out using various polarization combinations but a strong polarization dependence of the phenomena we studied (i.e. chirp asymmetry) was not noted. 

As a demonstration of the chirp asymmetry, we tuned the probe (D2) onto absorption by the $^{87}$Rb F=1 and then swept the pump (D1) through the Rb series.  By applying a triangle waveform to the laser diode driver (Thorlabs LD500) of the pump we then recorded on the fast detector a series of transparency spikes caused by populations changes as the pump was swept up and down across the $^{87}$Rb F=1 D1 resonances. We focussed on the $^{87}$Rb F=1 transitions since it had the largest excited state hyperfine splitting, allowing an unambigious D1 sweep through a single manifold of degenerate excited states. That is importantly not the case for the D2 probe there which always had overlapping, non-symmetric Doppler-broadened contributions to the transparency spikes. 

% set ylabel "D2 transmitted signal" font "computer modern, 32pt" offset -0.5
% set xlabel "Microseconds" font "computer modern, 32pt" offset 1
%  set lmargin -8
%  set bmargin 5
% unset key
% set datafile separator "," 
% q1 = 258
% set format x "%3.0f"
% set format y "%0.3f"
% set xtics font "computer modern, 24pt" scale 2
% set ytics font "computer modern, 24pt" scale 2
% plot [-30:20] "s13" using ($1*1e6-q1):2 w l, "s13" using  (.9*(-$1*1e6+30+q1)):2 w l
% set terminal postscript eps size 6.5,4.62 enhanced color font "Times Roman,20" linewidth 3
%  set output 'fig4a.eps'
% replot
%%%%%%%%%%% (all else the same but...) 
% set ylabel "D1 transmitted signal" font "computer modern, 32pt" offset -0.5
% plot [46:70] "SDS9_da_8_13_smoothed.txt" using ($1*1e6):2 w l lw 3
%  set output 'fig4b.eps'

A 200 MHz (1 GB/s sampling) deep storage 4-channel oscilloscope was used to simultaneously record the weak transmitted probe (FPD610-FS-NIR fast detector) and the bright pump beam (commercial <10MHz bandwidth). Each frequency sweep (of the pump) was of sufficient mode-hop free range to always cover multiple ground state $^{85}$Rb and $^{87}$Rb doppler-broadened resonances, all well resolved in the pump transmission trace. From these data we determined the sweep speed (we averaged the up- and down- chirp speeds which inside each trace differed from each other by less than 10 percent across this entire study). At intermediate sweep speeds a chirp asymmetry is evident by eye in the oscilloscope traces (Fig. \ref{fig:exampleTraces}). For simplicity we then  quantified the transmission maxima of the dominant excited state hyperfine line in the D2 probe during the pump (D1) sweeps in both the up- and down- chirps at various chirp speeds our measure of asymmetry throughout being (downmax-upmax)/(downmax+upmax) (see Fig. \ref{fig:asymExp}). 

\begin{figure} 
\includegraphics[width=2.4in]{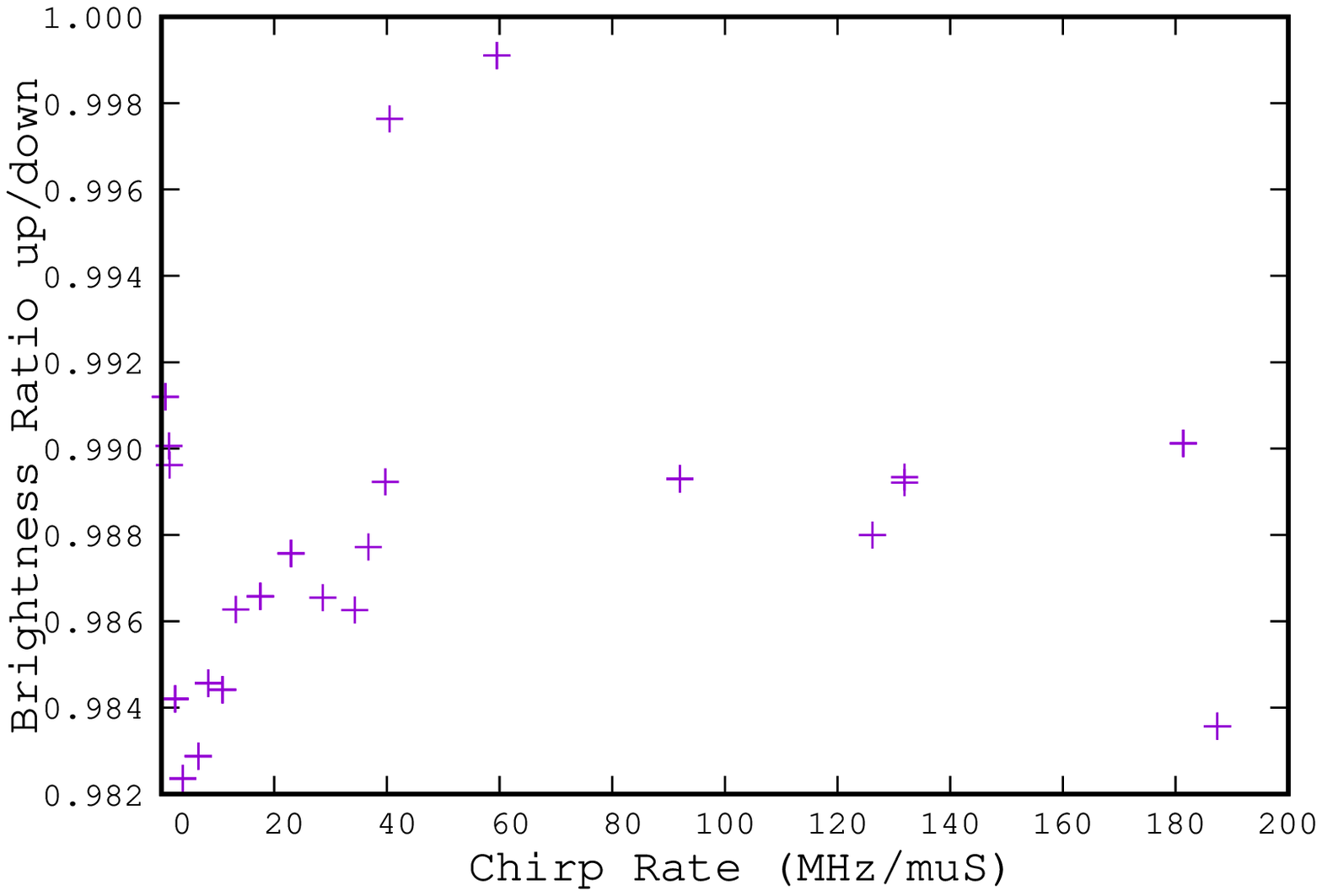}
\includegraphics[width=2.4in]{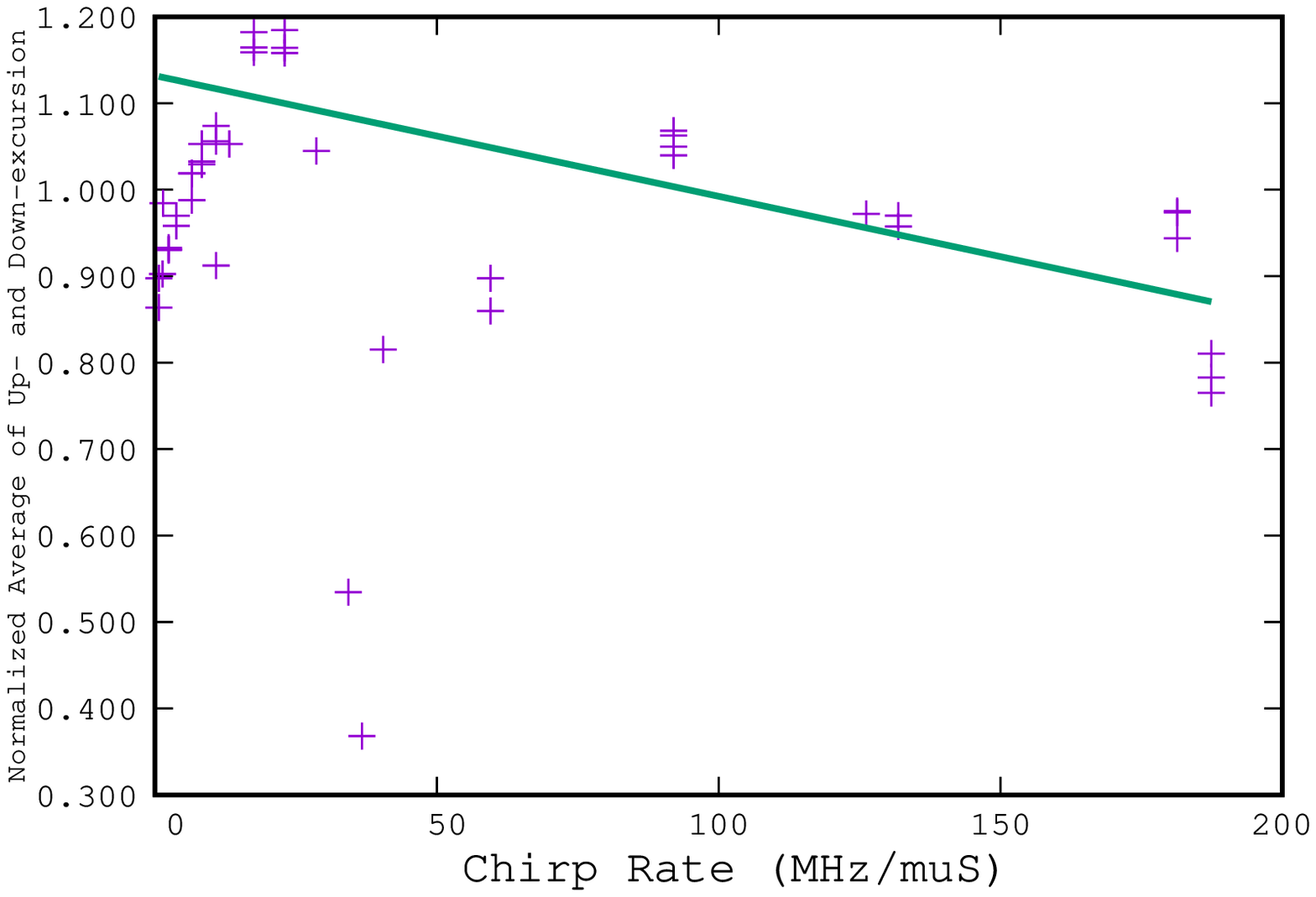}
\caption{(a) The experimentally measured atomic absorption ratio (up/down) in the D1 free-running laser "pump" light, if systematically different than one, would directly contribute to our measurement of the chirp asymmetry. (b) A plot of the average up- and down- D2  probe light transmission spikes amplitudes due to the D1 chirp, as a function of the latter's chirp speed. The green line is a linear fit to all the data after 15 MHz/$\mu$S. 
%A sufficiently large nonzero slope in these data could complicate the interpretation of the asymmetry we naively measure for each trace consequent to the laser diode driver's small systematic difference between up- chirp and down-chirp $\alpha$. However, 
The measured difference between intra-trace up- and down- chirp speeds was never more than 10 \%, indicating a resultant small systematic contribution to Fig.\ref{fig:asymExp} that we for simplicity ignore.} 
\label{fig:FPdata}
\end{figure}

Any significant memory effects in the laser's spectral lineshape or beam characteristics or filtering effects of the detection electronics could, in principle, complicate the interpretation of the observed chirp asymmetry as a consequence of the atomic response. Non-atomic contributions to the asymmetry may come from at least four sources. The pump laser's brightness may have systematically varied between up and down chirp, and furthermore that ratio of these intensities might vary with chirp speed. We quantify any 'brightness asymmetry' by determining the pump laser's brightness at the absorption peaks in the Rb vapor during each trace. Owing to the large spectral (and thus temporal in a sweep) width of the single-photon absorption doppler envelopes, that data stream has only lower frequency components, thus we use the pump data recorded on the slower commercial detector. Figure \ref{fig:FPdata} is a plot of the ratio (up-/down-chirp) of the laser intensity at the Rb resonance as a function of chirp speed. Although small, this effect has been removed from  the data presented in Fig. \ref{fig:asymExp}. 

As remarked earlier, the time constant of the laser diode current driver tended to make the average pump upchirp rates $|\alpha|$ a bit smaller than the average downchirp (less than 10 \% variation in each sweep across all these data). If the average probe transmission peak height varied precipitously enough with sweep speed this difference in chirp speed 
could contribute to the measured asymmetry. Again, this would complicate the interpretation of the intra-trace chirp asymmetry as dominated by effective $CP$-violation in the atomic response. Figure \ref{fig:FPdata}b is a graph of the measured normalized average measured probe transmission pulse amplitude as a function of pump chirp speed. Its relative flatness even with a systematic 10 \% $|\alpha|$ difference between up- and down- chirps,  certifies that intra-trace differences in $|\alpha|$ do not apparently materially contribute more than at most a few percent to the chirp asymmetry as measured experimentally and displayed in Fig.\ref{fig:asymExp}. 

A third potentially confounding non-atomic contribution to the asymmetry signal could come from systematic laser spectral changes with chirp speed. We therefore studied the pump beam at laser chirps of  118, 190 and 330 MHz/$\mu$S in transmission through a commercial Fabry-Perot (FP) cavity (Thorlabs SA200-5B, 540-820nm spectral range, FSR = 1500MHz, finesse $~$ 250) using , again, the  FPD610-FS-NIR as our detector. Fitting these data with skew lorentzians, the measured linewidth changes between up- and down- chirps were at most at the few percent level, as were the fit residuals themselves,  consistent with no systematic laser linewidth chirp asymmetry with these sweep speeds. 

Finally, the fact that the FP output as measured by the high speed amplified detector were unaffected by high sweep speeds indicated that there were negligible electronic bandwidth filtering effects in the entire detection chain.  

While the forgoing laser optical and electronics checks do not rule out non-atomic contributions to the measured chirp asymmetry, the magnitude of our observed asymmetry strongly suggests non-atomic effects were subdominant. We conclude that the experimentally observed up- and down- chirp asymmetry in two-color SAS is dominated by effects originating in the optical response of the atoms themselves.

% set ylabel "Brightness Ratio up/down" font "computer modern, 32pt" offset -0.5
% set xlabel "Chirp Rate (MHz/muS)" font "computer modern, 32pt" offset 1
%  set lmargin -8
%  set bmargin 5
% unset key
% set pointsize 3
% set format x "%3.0f"
% set format y "%0.3f"
% set xtics font "computer modern, 24pt" scale 2
% set ytics font "computer modern, 24pt" scale 2
% plot "data2.dat" using ($4/($2-$3)):($9/$11)
% set terminal postscript eps size 6.5,4.62 enhanced color font "Times Roman,20" linewidth 3
%  set output 'fig5a.eps'
% replot
%%%%%%%%%%%%%%% (for figure b, changes only) 
% f1(x) = a1*x+b1
% a1              = -8.62584e-05
% b1              = 0.0699849
% c1              = 0.0618313
% set pointsize 3
% set ylabel "Normalized Average of Up- and Down-excursion" font "computer modern, 20pt" offset -0.5
% plot [0:200] "data2.dat" using ($4/($2-$3)):(($6+$5-2.0*$7)/c1), "data2.dat"using ($4/($2-$3)):(f1(($4/($2-$3)))/c1) w l lw 3
%  set output 'fig5b.eps'

\section{Discussion}
\label{sec:results}
\begin{figure} 
\includegraphics[width=\linewidth]{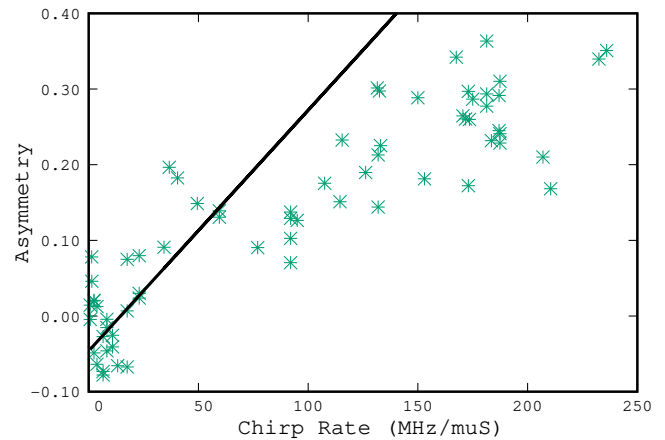}
\caption{Experimentally measured asymmetry with sweep rate. The black line is a linear least squares fit to the first third of the data, suggesting a linear increase followed by saturation at high chirp speeds. Going to yet larger chirp was not useful because of the onset of ringing as in Fig.\ref{fig:exampleTraces}b, behavior not possible in the basic leptogenesis Eq.~(\ref{eq:leptogenesis}) or optical population only models Eq.~(\ref{eq:populations2}), but a consequence of coherences as seen in high chirp numerical solutions of the semiclassical quantum optics model Eqs.~(\ref{eq:rho00}-\ref{eq:rho31}). 
}
\label{fig:asymExp}
\end{figure}

Since as remarked earlier the relaxation processes in the optical experiment always return the  optical signal to zero at large times/detunings, it is most straightforward at each chirp to compare the asymmetry of the maxima of the optical model with the maximum lepton number ({\it i.e.} prewashout maximum lepton number) in a leptogensis model having subsequent lepton decay. Experimentally one finds  the asymmetry of these peak values increases linearly with chirp speed and then saturates, as summarized in the experimental findings Fig.\ref{fig:asymExp}.

\begin{figure}
\includegraphics[width=2.5in]{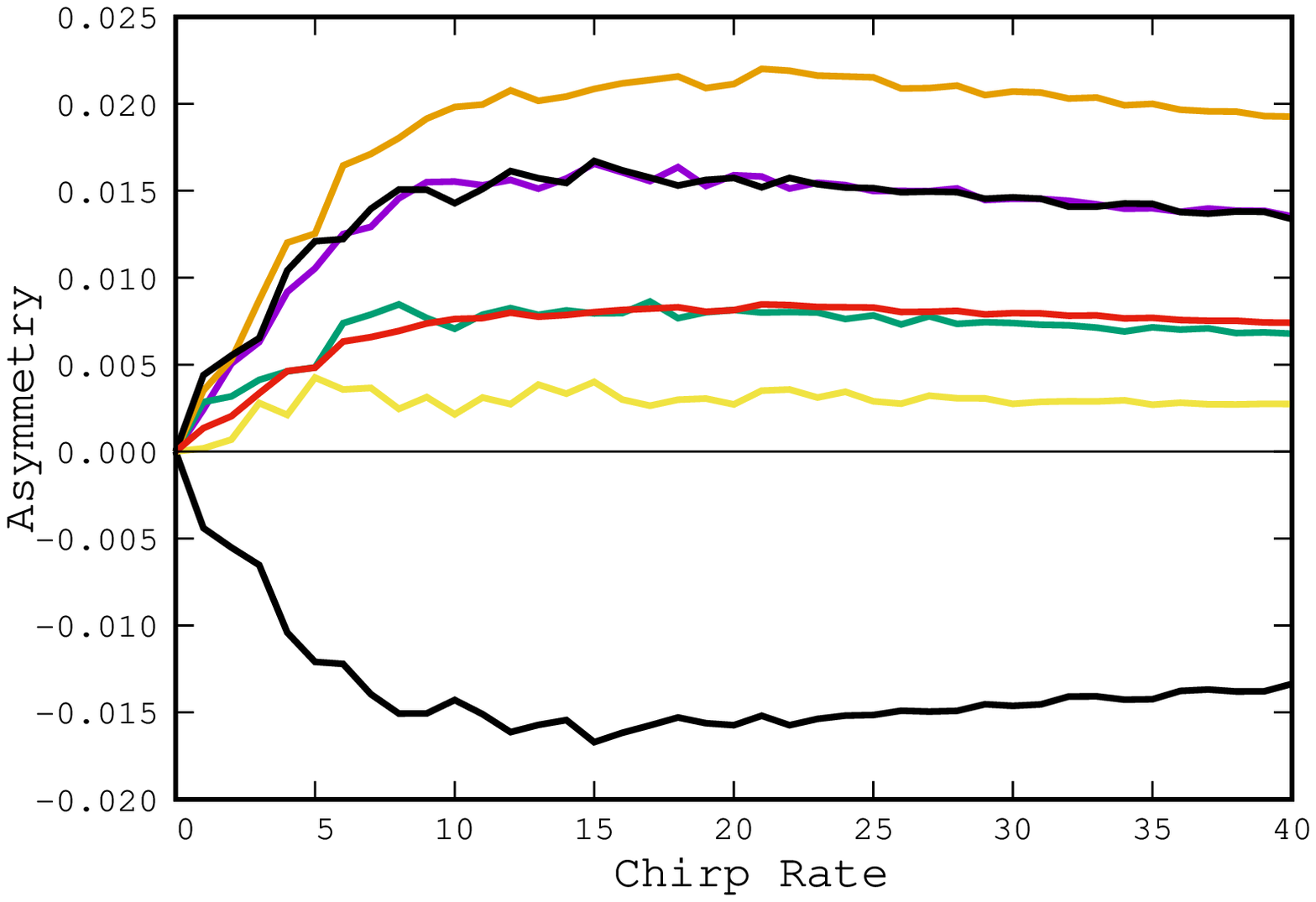}
\includegraphics[width=2.5in]{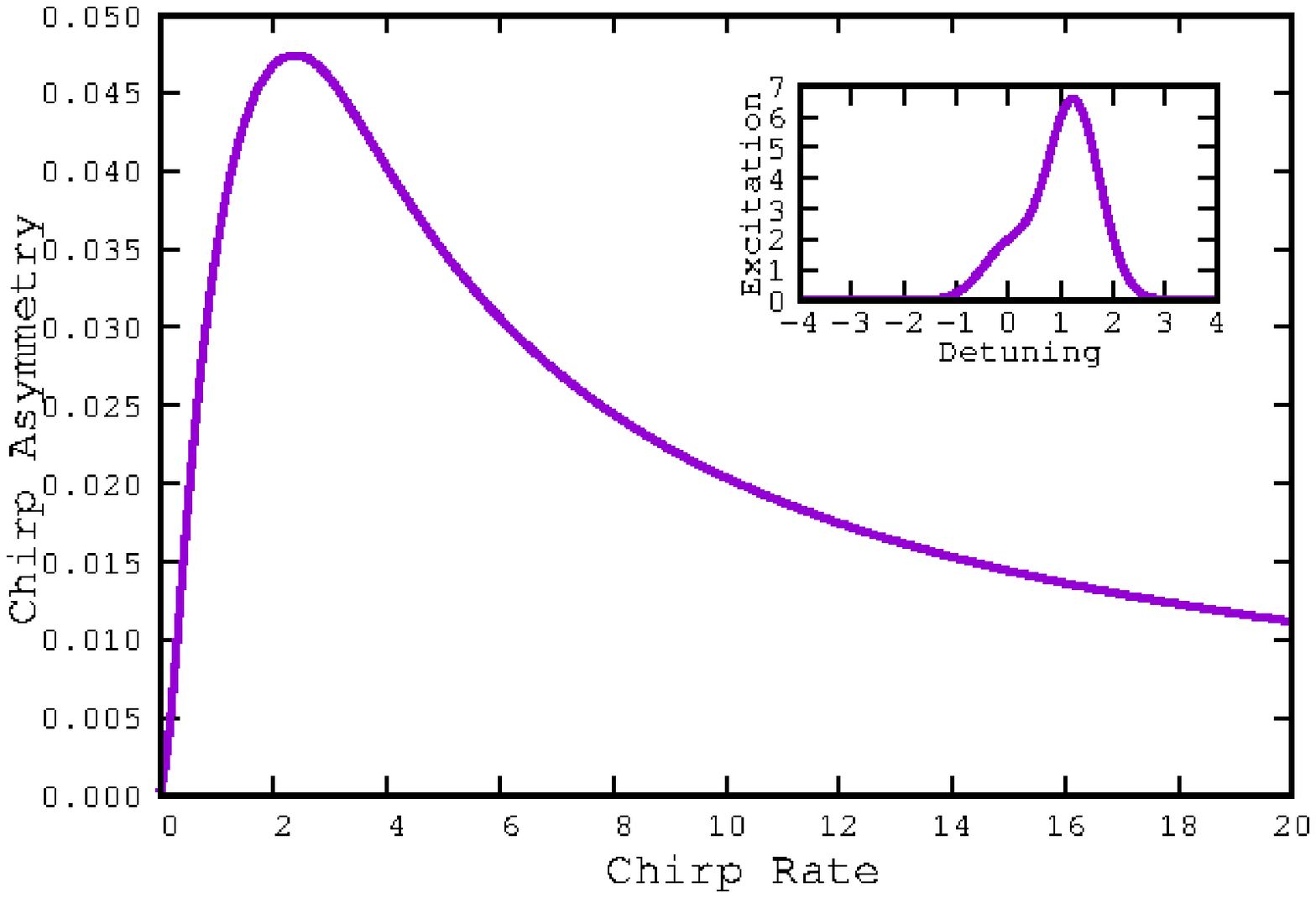}
\caption{(a) Asymmetry in semi-classical quantum optics four level model as in Eqs.~(\ref{eq:rho00}-\ref{eq:rho31}) with chirp speed. Curves are, in order, of $\delta' = -50 (black), 10, 25, 50, 65 (orange)$ $\gamma$ (Doppler width was 50$\gamma$). The black line has been plotted along with its negative showing symmetry of the effect in the sign of $\delta'$. The red line is the asymmetry of the $\delta'=65$ line scaled by 25/65, showing the asymmetry's approximate linearity in $\delta'$. 
(b) Asymmetry maxima in basic two-level population model as described in Eq.~(\ref{eq:populations}) as a function of the chirp speed. The linear rise at slow chirp speed plateaus and decreases at high chirp speed due to the fact that in these simulations there is a fixed timescale for the production of what serves as the optical analogue of the 'heavy neutrino' in the leptogenesis model; if the chirp rate is too fast then fewer of those states are created in the early stages of the chirp. Inset shows the intrinsic asymmetric lineshape used in that evaluation of the two-level population model} 
\label{fig:asymTheory}
\end{figure}

\begin{figure}
\includegraphics[width=\linewidth]{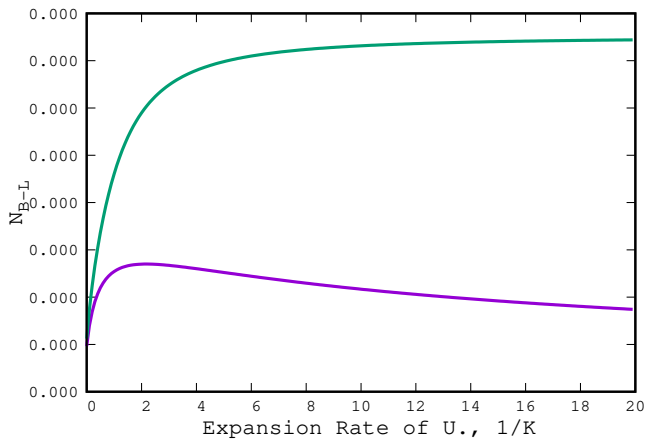}
\caption{Basic leptogenesis theory model output for net lepton number as a function of the hubble rate. Saturation at high sweep rates results from the product of the limited number of heavy leptons made during that thermal epoch that subsequently freeze out for a quick expanding/cooling universe and the intrinsic $C$-violating lepton number creating parameter $\epsilon$.} 
\label{fig:asymLeptogenesis}
\end{figure}

% set ylabel "Asymmetry" font "computer modern, 32pt" offset -0.5
% set xlabel "Chirp Rate (MHz/muS)" font "computer modern, 32pt" offset 1
%  set lmargin -8
%  set bmargin 5
% unset key
% set pointsize 3
% set format x "%3.0f"
% set format y "%0.2f"
% set xtics font "computer modern, 24pt" scale 2
% set ytics font "computer modern, 24pt" scale 2
% f1(x) = a1*x+b1
% a1              = 0.00318086
% b1              = -0.04612
%  plot [0:250] [-0.1:0.4] "21data" using (($2+$3)/2.0):(($5-$4)*2.0/($4+$5-2.0*$6)) pt 3 lc 2, "data.dat" using ($4/($2-$3)):((($6-$7)/$10-($5-$7)/$8)/($6/$10+$5/$8-$7*(1.0/$10+1.0/$8))*2.0) pt 3 lc 2, "data.dat" using ($4/($2-$3)):(f1(($4/($2-$3)))) w l lw 3 lc 8
% set terminal postscript eps size 6.5,4.62 enhanced color font "Times Roman,20" linewidth 3
%  set output 'fig6.eps'
% replot
This is corroborated by the evaluation of the semiclassical quantum optics model as described above for our two-color SAS model in which we have included explicit $CP$ violation by including an asymmetric density of states via a nonzero, fixed probe detuning $\delta'$. Fig.\ref{fig:asymTheory}a clearly displays a linear regime at slow chirp speeds again saturating at high chirp speed. Studies are underway to determine how the chirp speed at the onset of the limiting asymmetry is related to accessible experimental parameters {\it i.e.}  probe detuning and pump intensity. We note in Fig.\ref{fig:asymLeptogenesis} the strong phenomenological similarity of the above with the chirp speed dependence of the maximum lepton number in the basic leptogenesis model of Eq.~(\ref{eq:leptogenesis}), as well as that in the naive population model Fig.\ref{fig:asymTheory}b. To re-itterate, in both Fig.\ref{fig:asymExp} and Fig.\ref{fig:asymTheory} the asymmetry plotted was the difference of the peak maxima in the down chirp minus that in the up chirp all over the sum. 

The sign of the measured asymmetry is also consistent with the spectral asymmetry. Note that in the evaluation of the four-level quantum optics model for the system, for a detuning $\delta'=+60$ the downchirp of the pump occasions the larger peak probe response. Due to the finite spectral width of the laser and the Boltzmann averaging in that simulation, for $\delta'>0$  that single state $\ket{3}$ results in the density of probe-active atoms that at first large and then decreases during the pump downchirp through resonance. 

This is consistent with the experiment in which, for $^{87}$Rb F=1 D2 transitions the spectral density (due to the excited hyperfine spectrum) is skewed towards higher energy. Thus, for a given fixed probe frequency near zero in the experiment, the density of slightly red detuned atoms is larger than that of blue detuned ones, as in the $\delta'>0$ case for the state $\ket{3}$ quantum optics model with finite laser linewidth. Thus experimentally we expect that the optical response to the probe to be larger for the pump downchirp, resulting in a positive asymmetry (Fig.\ref{fig:asymExp}). 
 
Finally note that in this case, since due to the memory effects in the laser diode driver electronics, our downchirps corresponded to sweep rates on average a bit more than for the upchirps the small negative slope of the trend line of Fig.\ref{fig:FPdata}b would tend to have reduced our measured asymmetry 
Fig.\ref{fig:asymExp}. 

% set ylabel "N_B_-_L" font "computer modern, 32pt" offset -0.7
% set xlabel "Expansion Rate of U., 1/K" font "computer modern, 32pt" offset 1
%  set lmargin -8
%  set bmargin 5
% set format x "%2.0f"
% set format y "%0.3f"
% set xtics font "computer modern, 24pt" scale 2
% set ytics font "computer modern, 24pt" scale 2
% unset key
%plot  "lambda_is_half" using ($0/10):1 w l lw 3, "lambda_is_zero" using ($0/10):1 w l lw 3
% set terminal postscript eps size 6.5,4.62 enhanced color font "Times Roman,20" linewidth 3
%  set output 'fig8.eps'
% replot

% Section on Chemistry tie-in
Our observations are summarized as a straightforward generalization of the Sakharov conditions: a system with a broken discrete symmetry that starts in a symmetrical state will result in a degree of asymmetry (up to a limit) consonant with how far off equilibrium it has been driven. Chirp speed is a convenient proxy for the later. To cite one example not related to broken conjugation invariance, consider some modern approaches to enantomeric resolutions in chiral chemistry. There parity is the broken discrete symmetry and a non-equilibrium process (such as the grinding of crystals in, for example, Viedma ripening) drives the growth of the enantomeric excess\cite{frank, viedma, Breveglieri}. Further, as indicated in Figure 6 of Ref. \cite{Breveglieri} quantifying the per-cycle change in the enantomeric excess as a function of the temperature cycling (here cooling rate, the analogue to the "chirp rate") rate indicates a monotonic dependence that saturates. The final enantomeric excess in this process can be of either species, so Viedma ripening is an example of spontaneous effective $CP$-breaking under chirp.

These deliberations also elucidate up- and down-chirp asymmetry in pulsed two color transitions. Ref.\cite{broers} described a two color transition in which the optical design allowed broad control over the pulse sequence. That system consisted of three participating states in which the ground state and the middle state were connected by one of the colors and the middle state and the target state were connected by the other. Initially only the ground state is occupied. In that reference it was noted that  the so-called "non-intuitive pulse sequence" of illuminating the transition coupling the unoccupied middle and target states before exciting the ground and middle state's transition can result in a larger population transfer than the "intuitive sequence" of the other ordering. In either a population model or the full quantum optics model to first order (i.e. no AC Stark effect or other non-linear effect beyond two-level saturation) for this system the equations themselves are effective $CP$-symmetric, indicating that time reversing the pulse  sequence would be expected to give the same population transfer. However, it is straightforward to see that including the leading AC Stark effect as an intensity dependent detuning in this quantum optics model breaks the conjugation symmetry and then leads to an asymmetry between each up- and down-chirped  two color pulse sequence. We refer to AC Stark induced breaking of the discrete symmetry in that case as induced effective $CP$-breaking in the sense that it is modulated by a field endemic to the process whereas in the rest of this note we've assumed explicit $CP$-breaking in that the spectral density of the transition is not symmetric in the detuning. This conclusion is independent of the lifetime of the middle state.
To re-iterate, whatever the source of the $CP$-breaking, it is in concert with the chirp speed that time evolution leads to final state asymmetry, displaying the unifying picture of the Sakharov conditions.

We have developed and tested an approach to understanding the effect of a broken discrete symmetry in a chirped optical process, and showed that the generic features include initially linear growth of the asymmetry which saturates at higher chirp speeds. As an optical spectroscopic embodiment of this idea we studied how the observed lineshape asymmetry in two-color saturated absorption spectroscopy changes with chirp speed of the pump. The sign, growth and saturation of the optical asymmetry between up- and down- chirp is quantitatively described by the basic quantum mechanical analysis of the atom-light interaction. This portion of optical theory and the experimental phenomenology can be straightforwardly 'mapped' onto a more general framework that satisfies the Sakharov conditions for leptogenesis. We expect the observed phenomenology described in this article to be highly conserved across disparate systems satisfying the Sakharov conditions as evidenced in earlier literature on chirality selection in a chemical system driven out of equilibrium and separately in a two-color optical pulsed excitation processes. 

\section*{Funding}
Crescimanno received partial support from NSF grant (DMR-1609077) during the initial stages of this study. Commons, Jones, George and Weiser were supported by the Ann Seimon endowment for undergraduate research at YSU. 

\section*{Acknowledgements}
We are grateful for discussions with David F. Phillips, Nuria Rius, Al Shapere, Doug Genna and Peter Norris. 

\section*{Disclosure} 
The authors declare no conflicts of interest.

\bibliography{refs}

\begin{thebibliography}{60}%
\makeatletter
\providecommand \@ifxundefined [1]{%
 \@ifx{#1\undefined}
}%
\providecommand \@ifnum [1]{%
 \ifnum #1\expandafter \@firstoftwo
 \else \expandafter \@secondoftwo
 \fi
}%
\providecommand \@ifx [1]{%
 \ifx #1\expandafter \@firstoftwo
 \else \expandafter \@secondoftwo
 \fi
}%
\providecommand \natexlab [1]{#1}%
\providecommand \enquote  [1]{``#1''}%
\providecommand \bibnamefont  [1]{#1}%
\providecommand \bibfnamefont [1]{#1}%
\providecommand \citenamefont [1]{#1}%
\providecommand \href@noop [0]{\@secondoftwo}%
\providecommand \href [0]{\begingroup \@sanitize@url \@href}%
\providecommand \@href[1]{\@@startlink{#1}\@@href}%
\providecommand \@@href[1]{\endgroup#1\@@endlink}%
\providecommand \@sanitize@url [0]{\catcode `\\12\catcode `\$12\catcode
  `\&12\catcode `\#12\catcode `\^12\catcode `\_12\catcode `\%12\relax}%
\providecommand \@@startlink[1]{}%
\providecommand \@@endlink[0]{}%
\providecommand \url  [0]{\begingroup\@sanitize@url \@url }%
\providecommand \@url [1]{\endgroup\@href {#1}{\urlprefix }}%
\providecommand \urlprefix  [0]{URL }%
\providecommand \Eprint [0]{\href }%
\providecommand \doibase [0]{http://dx.doi.org/}%
\providecommand \selectlanguage [0]{\@gobble}%
\providecommand \bibinfo  [0]{\@secondoftwo}%
\providecommand \bibfield  [0]{\@secondoftwo}%
\providecommand \translation [1]{[#1]}%
\providecommand \BibitemOpen [0]{}%
\providecommand \bibitemStop [0]{}%
\providecommand \bibitemNoStop [0]{.\EOS\space}%
\providecommand \EOS [0]{\spacefactor3000\relax}%
\providecommand \BibitemShut  [1]{\csname bibitem#1\endcsname}%
\let\auto@bib@innerbib\@empty
%</preamble>
\bibitem [{\citenamefont {Viedma}(2005)}]{viedma}%
  \BibitemOpen
  \bibfield  {author} {\bibinfo {author} {\bibfnamefont {C.}~\bibnamefont
  {Viedma}},\ }\href@noop {} {\bibfield  {journal} {\bibinfo  {journal} {Phys.
  Rev. Lett.}\ }\textbf {\bibinfo {volume} {94}},\ \bibinfo {pages} {065504}
  (\bibinfo {year} {2005})}\BibitemShut {NoStop}%
\bibitem [{\citenamefont {Broers}\ \emph {et~al.}(1992)\citenamefont {Broers},
  \citenamefont {Heuvell},\ and\ \citenamefont {Noordam}}]{broers}%
  \BibitemOpen
  \bibfield  {author} {\bibinfo {author} {\bibfnamefont {B.}~\bibnamefont
  {Broers}}, \bibinfo {author} {\bibfnamefont {H.~B. V. L. V.~D.}\ \bibnamefont
  {Heuvell}}, \ and\ \bibinfo {author} {\bibfnamefont {L.~D.}\ \bibnamefont
  {Noordam}},\ }\href {\doibase 10.1103/physrevlett.69.2062} {\bibfield
  {journal} {\bibinfo  {journal} {Physical Review Letters}\ }\textbf {\bibinfo
  {volume} {69}},\ \bibinfo {pages} {2062–2065} (\bibinfo {year}
  {1992})}\BibitemShut {NoStop}%
\bibitem [{\citenamefont {Sakharov}(1991)}]{sakharov91}%
  \BibitemOpen
  \bibfield  {author} {\bibinfo {author} {\bibfnamefont {A.~D.}\ \bibnamefont
  {Sakharov}},\ }\href {\doibase 10.1070/pu1991v034n05abeh002497} {\bibfield
  {journal} {\bibinfo  {journal} {Soviet Physics Uspekhi}\ }\textbf {\bibinfo
  {volume} {34}},\ \bibinfo {pages} {392} (\bibinfo {year} {1991})}\BibitemShut
  {NoStop}%
\bibitem [{\citenamefont {Buchmüller}\ \emph {et~al.}(2005)\citenamefont
  {Buchmüller}, \citenamefont {Di~Bari},\ and\ \citenamefont
  {Pl\"umacher}}]{buchmuller1}%
  \BibitemOpen
  \bibfield  {author} {\bibinfo {author} {\bibfnamefont {W.}~\bibnamefont
  {Buchmüller}}, \bibinfo {author} {\bibfnamefont {P.}~\bibnamefont
  {Di~Bari}}, \ and\ \bibinfo {author} {\bibfnamefont {M.}~\bibnamefont
  {Pl\"umacher}},\ }\href@noop {} {\bibfield  {journal} {\bibinfo  {journal}
  {Annals of Physics}\ }\textbf {\bibinfo {volume} {315}},\ \bibinfo {pages}
  {305} (\bibinfo {year} {2005})}\BibitemShut {NoStop}%
\bibitem [{\citenamefont {Buchm\"uller}\ \emph {et~al.}(2005)\citenamefont
  {Buchm\"uller}, \citenamefont {Peccei},\ and\ \citenamefont
  {Yanagida}}]{buchmuller2}%
  \BibitemOpen
  \bibfield  {author} {\bibinfo {author} {\bibfnamefont {W.}~\bibnamefont
  {Buchm\"uller}}, \bibinfo {author} {\bibfnamefont {R.~D.}\ \bibnamefont
  {Peccei}}, \ and\ \bibinfo {author} {\bibfnamefont {T.}~\bibnamefont
  {Yanagida}},\ }\href@noop {} {\bibfield  {journal} {\bibinfo  {journal}
  {Annu. Rev. Nucl. Part. Sci.}\ }\textbf {\bibinfo {volume} {55}},\ \bibinfo
  {pages} {11} (\bibinfo {year} {2005})}\BibitemShut {NoStop}%
\bibitem [{\citenamefont {Gavela}\ \emph {et~al.}(1994)\citenamefont {Gavela},
  \citenamefont {Hernandez}, \citenamefont {Orloff},\ and\ \citenamefont
  {Pene}}]{GHOP}%
  \BibitemOpen
  \bibfield  {author} {\bibinfo {author} {\bibfnamefont {M.~B.}\ \bibnamefont
  {Gavela}}, \bibinfo {author} {\bibfnamefont {P.}~\bibnamefont {Hernandez}},
  \bibinfo {author} {\bibfnamefont {J.}~\bibnamefont {Orloff}}, \ and\ \bibinfo
  {author} {\bibfnamefont {O.}~\bibnamefont {Pene}},\ }\href@noop {} {\bibfield
   {journal} {\bibinfo  {journal} {Modern Physics Letters A}\ }\textbf
  {\bibinfo {volume} {09}},\ \bibinfo {pages} {795} (\bibinfo {year}
  {1994})}\BibitemShut {NoStop}%
\bibitem [{\citenamefont {Kuzmin}\ \emph {et~al.}(1985)\citenamefont {Kuzmin},
  \citenamefont {Rubakov},\ and\ \citenamefont {Shaposhnikov}}]{kuzmin}%
  \BibitemOpen
  \bibfield  {author} {\bibinfo {author} {\bibfnamefont {V.}~\bibnamefont
  {Kuzmin}}, \bibinfo {author} {\bibfnamefont {V.}~\bibnamefont {Rubakov}}, \
  and\ \bibinfo {author} {\bibfnamefont {M.}~\bibnamefont {Shaposhnikov}},\
  }\href {\doibase 10.1016/0370-2693(85)91028-7} {\bibfield  {journal}
  {\bibinfo  {journal} {Physics Letters B}\ }\textbf {\bibinfo {volume}
  {155}},\ \bibinfo {pages} {36–42} (\bibinfo {year} {1985})}\BibitemShut
  {NoStop}%
\bibitem [{\citenamefont {Huet}\ and\ \citenamefont {Sather}(1995)}]{huet}%
  \BibitemOpen
  \bibfield  {author} {\bibinfo {author} {\bibfnamefont {P.}~\bibnamefont
  {Huet}}\ and\ \bibinfo {author} {\bibfnamefont {E.}~\bibnamefont {Sather}},\
  }\href@noop {} {\bibfield  {journal} {\bibinfo  {journal} {Physical Review
  D}\ }\textbf {\bibinfo {volume} {51}},\ \bibinfo {pages} {379} (\bibinfo
  {year} {1995})}\BibitemShut {NoStop}%
\bibitem [{\citenamefont {Kajantie}\ \emph {et~al.}(1996)\citenamefont
  {Kajantie}, \citenamefont {Laine},\ and\ \citenamefont
  {Shaposhnikov}}]{kajantie}%
  \BibitemOpen
  \bibfield  {author} {\bibinfo {author} {\bibfnamefont {K.}~\bibnamefont
  {Kajantie}}, \bibinfo {author} {\bibfnamefont {K.}~\bibnamefont {Laine},
  \bibfnamefont {M.~Rummukainen}}, \ and\ \bibinfo {author} {\bibfnamefont
  {M.}~\bibnamefont {Shaposhnikov}},\ }\href@noop {} {\bibfield  {journal}
  {\bibinfo  {journal} {Physical Review Letters}\ }\textbf {\bibinfo {volume}
  {77}},\ \bibinfo {pages} {2887} (\bibinfo {year} {1996})}\BibitemShut
  {NoStop}%
\bibitem [{\citenamefont {Blanchet}\ \emph {et~al.}(2013)\citenamefont
  {Blanchet}, \citenamefont {Bari}, \citenamefont {Jones},\ and\ \citenamefont
  {Marzola}}]{blanchet}%
  \BibitemOpen
  \bibfield  {author} {\bibinfo {author} {\bibfnamefont {S.}~\bibnamefont
  {Blanchet}}, \bibinfo {author} {\bibfnamefont {P.~D.}\ \bibnamefont {Bari}},
  \bibinfo {author} {\bibfnamefont {D.~A.}\ \bibnamefont {Jones}}, \ and\
  \bibinfo {author} {\bibfnamefont {L.}~\bibnamefont {Marzola}},\ }\href
  {\doibase 10.1088/1475-7516/2013/01/041} {\bibfield  {journal} {\bibinfo
  {journal} {Journal of Cosmology and Astroparticle Physics}\ }\textbf
  {\bibinfo {volume} {2013}},\ \bibinfo {pages} {041} (\bibinfo {year}
  {2013})}\BibitemShut {NoStop}%
\bibitem [{\citenamefont {Mangano}\ and\ \citenamefont
  {Miele}(2000)}]{mangano}%
  \BibitemOpen
  \bibfield  {author} {\bibinfo {author} {\bibfnamefont {G.}~\bibnamefont
  {Mangano}}\ and\ \bibinfo {author} {\bibfnamefont {G.}~\bibnamefont
  {Miele}},\ }\href {\doibase 10.1103/PhysRevD.62.063514} {\bibfield  {journal}
  {\bibinfo  {journal} {Phys. Rev. D}\ }\textbf {\bibinfo {volume} {62}},\
  \bibinfo {pages} {063514} (\bibinfo {year} {2000})}\BibitemShut {NoStop}%
\bibitem [{\citenamefont {Avila}\ \emph {et~al.}(1987)\citenamefont {Avila},
  \citenamefont {Giordano}, \citenamefont {Candelier}, \citenamefont {Clercq},
  \citenamefont {Theobald},\ and\ \citenamefont {Cerez}}]{avila}%
  \BibitemOpen
  \bibfield  {author} {\bibinfo {author} {\bibfnamefont {G.}~\bibnamefont
  {Avila}}, \bibinfo {author} {\bibfnamefont {V.}~\bibnamefont {Giordano}},
  \bibinfo {author} {\bibfnamefont {V.}~\bibnamefont {Candelier}}, \bibinfo
  {author} {\bibfnamefont {E.~D.}\ \bibnamefont {Clercq}}, \bibinfo {author}
  {\bibfnamefont {G.}~\bibnamefont {Theobald}}, \ and\ \bibinfo {author}
  {\bibfnamefont {P.}~\bibnamefont {Cerez}},\ }\href {\doibase
  10.1103/physreva.36.3719} {\bibfield  {journal} {\bibinfo  {journal}
  {Physical Review A}\ }\textbf {\bibinfo {volume} {36}},\ \bibinfo {pages}
  {3719} (\bibinfo {year} {1987})}\BibitemShut {NoStop}%
\bibitem [{\citenamefont {Gunaratne}\ \emph {et~al.}(2007)\citenamefont
  {Gunaratne}, \citenamefont {Zhu}, \citenamefont {Lozovoy},\ and\
  \citenamefont {Dantus}}]{gunaratne}%
  \BibitemOpen
  \bibfield  {author} {\bibinfo {author} {\bibfnamefont {T.~C.}\ \bibnamefont
  {Gunaratne}}, \bibinfo {author} {\bibfnamefont {X.}~\bibnamefont {Zhu}},
  \bibinfo {author} {\bibfnamefont {V.~V.}\ \bibnamefont {Lozovoy}}, \ and\
  \bibinfo {author} {\bibfnamefont {M.}~\bibnamefont {Dantus}},\ }\href
  {\doibase 10.1016/j.chemphys.2007.04.027} {\bibfield  {journal} {\bibinfo
  {journal} {Chemical Physics}\ }\textbf {\bibinfo {volume} {338}},\ \bibinfo
  {pages} {259} (\bibinfo {year} {2007})}\BibitemShut {NoStop}%
\bibitem [{\citenamefont {Cardman}\ and\ \citenamefont
  {Raithel}(2020)}]{cardman}%
  \BibitemOpen
  \bibfield  {author} {\bibinfo {author} {\bibfnamefont {R.}~\bibnamefont
  {Cardman}}\ and\ \bibinfo {author} {\bibfnamefont {G.}~\bibnamefont
  {Raithel}},\ }\href {\doibase 10.1103/physreva.101.013434} {\bibfield
  {journal} {\bibinfo  {journal} {Physical Review A}\ }\textbf {\bibinfo
  {volume} {101}},\ \bibinfo {pages} {013434} (\bibinfo {year}
  {2020})}\BibitemShut {NoStop}%
\bibitem [{\citenamefont {Carroll}\ and\ \citenamefont
  {Hioe}(1986)}]{carroll1}%
  \BibitemOpen
  \bibfield  {author} {\bibinfo {author} {\bibfnamefont {C.~E.}\ \bibnamefont
  {Carroll}}\ and\ \bibinfo {author} {\bibfnamefont {F.~T.}\ \bibnamefont
  {Hioe}},\ }\href {\doibase 10.1088/0305-4470/19/17/022} {\bibfield  {journal}
  {\bibinfo  {journal} {Journal of Physics A: Mathematical and General}\
  }\textbf {\bibinfo {volume} {19}},\ \bibinfo {pages} {3579} (\bibinfo {year}
  {1986})}\BibitemShut {NoStop}%
\bibitem [{\citenamefont {Carroll}\ and\ \citenamefont
  {Hioe}(1992)}]{carroll2}%
  \BibitemOpen
  \bibfield  {author} {\bibinfo {author} {\bibfnamefont {C.~E.}\ \bibnamefont
  {Carroll}}\ and\ \bibinfo {author} {\bibfnamefont {F.~T.}\ \bibnamefont
  {Hioe}},\ }\href@noop {} {\bibfield  {journal} {\bibinfo  {journal} {Phys Rev
  Lett.}\ }\textbf {\bibinfo {volume} {68}},\ \bibinfo {pages} {3523} (\bibinfo
  {year} {1992})}\BibitemShut {NoStop}%
\bibitem [{\citenamefont {Paspalakis}\ \emph {et~al.}(1997)\citenamefont
  {Paspalakis}, \citenamefont {Protopapas},\ and\ \citenamefont
  {Knight}}]{Paspalakis}%
  \BibitemOpen
  \bibfield  {author} {\bibinfo {author} {\bibfnamefont {E.}~\bibnamefont
  {Paspalakis}}, \bibinfo {author} {\bibfnamefont {M.}~\bibnamefont
  {Protopapas}}, \ and\ \bibinfo {author} {\bibfnamefont {P.~L.}\ \bibnamefont
  {Knight}},\ }\href {\doibase https://doi.org/10.1016/S0030-4018(97)00254-X}
  {\bibfield  {journal} {\bibinfo  {journal} {Optics Communications}\ }\textbf
  {\bibinfo {volume} {142}},\ \bibinfo {pages} {34} (\bibinfo {year}
  {1997})}\BibitemShut {NoStop}%
\bibitem [{\citenamefont {Mcculloch}\ \emph {et~al.}(2006)\citenamefont
  {Mcculloch}, \citenamefont {Duxbury},\ and\ \citenamefont
  {Langford}}]{mcculloch}%
  \BibitemOpen
  \bibfield  {author} {\bibinfo {author} {\bibfnamefont {M.~T.}\ \bibnamefont
  {Mcculloch}}, \bibinfo {author} {\bibfnamefont {G.}~\bibnamefont {Duxbury}},
  \ and\ \bibinfo {author} {\bibfnamefont {N.}~\bibnamefont {Langford}},\
  }\href {\doibase 10.1080/00268970600857651} {\bibfield  {journal} {\bibinfo
  {journal} {Molecular Physics}\ }\textbf {\bibinfo {volume} {104}},\ \bibinfo
  {pages} {2767–2779} (\bibinfo {year} {2006})}\BibitemShut {NoStop}%
\bibitem [{\citenamefont {Shwa}\ and\ \citenamefont {Katz}(2014)}]{shwa}%
  \BibitemOpen
  \bibfield  {author} {\bibinfo {author} {\bibfnamefont {D.}~\bibnamefont
  {Shwa}}\ and\ \bibinfo {author} {\bibfnamefont {N.}~\bibnamefont {Katz}},\
  }\href {\doibase 10.1103/PhysRevA.90.023858} {\bibfield  {journal} {\bibinfo
  {journal} {Phys. Rev. A}\ }\textbf {\bibinfo {volume} {90}},\ \bibinfo
  {pages} {023858} (\bibinfo {year} {2014})}\BibitemShut {NoStop}%
\bibitem [{\citenamefont {Melinger}\ \emph {et~al.}(1995)\citenamefont
  {Melinger}, \citenamefont {Gandhi}, \citenamefont {Hariharan}, \citenamefont
  {Goswami},\ and\ \citenamefont {Warren}}]{melinger}%
  \BibitemOpen
  \bibfield  {author} {\bibinfo {author} {\bibfnamefont {J.~S.}\ \bibnamefont
  {Melinger}}, \bibinfo {author} {\bibfnamefont {S.~R.}\ \bibnamefont
  {Gandhi}}, \bibinfo {author} {\bibfnamefont {A.}~\bibnamefont {Hariharan}},
  \bibinfo {author} {\bibfnamefont {D.}~\bibnamefont {Goswami}}, \ and\
  \bibinfo {author} {\bibfnamefont {W.~S.}\ \bibnamefont {Warren}},\ }\href
  {\doibase 10.1063/1.469288} {\bibfield  {journal} {\bibinfo  {journal} {The
  Journal of Chemical Physics}\ }\textbf {\bibinfo {volume} {102}},\ \bibinfo
  {pages} {5574–5574} (\bibinfo {year} {1995})}\BibitemShut {NoStop}%
\bibitem [{\citenamefont {Liedenbaum}\ \emph {et~al.}(1989)\citenamefont
  {Liedenbaum}, \citenamefont {Stolte},\ and\ \citenamefont
  {Reuss}}]{liedenbaum}%
  \BibitemOpen
  \bibfield  {author} {\bibinfo {author} {\bibfnamefont {C.}~\bibnamefont
  {Liedenbaum}}, \bibinfo {author} {\bibfnamefont {S.}~\bibnamefont {Stolte}},
  \ and\ \bibinfo {author} {\bibfnamefont {J.}~\bibnamefont {Reuss}},\ }\href
  {\doibase 10.1016/0370-1573(89)90018-5} {\bibfield  {journal} {\bibinfo
  {journal} {Physics Reports}\ }\textbf {\bibinfo {volume} {178}},\ \bibinfo
  {pages} {1–24} (\bibinfo {year} {1989})}\BibitemShut {NoStop}%
\bibitem [{\citenamefont {Noel}\ \emph {et~al.}(2012)\citenamefont {Noel},
  \citenamefont {Dietrich}, \citenamefont {Kurz}, \citenamefont {Shu},
  \citenamefont {Wright},\ and\ \citenamefont {Blinov}}]{noel}%
  \BibitemOpen
  \bibfield  {author} {\bibinfo {author} {\bibfnamefont {T.}~\bibnamefont
  {Noel}}, \bibinfo {author} {\bibfnamefont {M.~R.}\ \bibnamefont {Dietrich}},
  \bibinfo {author} {\bibfnamefont {N.}~\bibnamefont {Kurz}}, \bibinfo {author}
  {\bibfnamefont {G.}~\bibnamefont {Shu}}, \bibinfo {author} {\bibfnamefont
  {J.}~\bibnamefont {Wright}}, \ and\ \bibinfo {author} {\bibfnamefont {B.~B.}\
  \bibnamefont {Blinov}},\ }\href {\doibase 10.1103/physreva.85.023401}
  {\bibfield  {journal} {\bibinfo  {journal} {Physical Review A}\ }\textbf
  {\bibinfo {volume} {85}},\ \bibinfo {pages} {023401} (\bibinfo {year}
  {2012})}\BibitemShut {NoStop}%
\bibitem [{\citenamefont {Zhou}\ \emph {et~al.}(2020)\citenamefont {Zhou},
  \citenamefont {Liu}, \citenamefont {Sun}, \citenamefont {An}, \citenamefont
  {Li}, \citenamefont {Bao},\ and\ \citenamefont {Pan}}]{zhou}%
  \BibitemOpen
  \bibfield  {author} {\bibinfo {author} {\bibfnamefont {M.-T.}\ \bibnamefont
  {Zhou}}, \bibinfo {author} {\bibfnamefont {J.-L.}\ \bibnamefont {Liu}},
  \bibinfo {author} {\bibfnamefont {P.-F.}\ \bibnamefont {Sun}}, \bibinfo
  {author} {\bibfnamefont {Z.-Y.}\ \bibnamefont {An}}, \bibinfo {author}
  {\bibfnamefont {J.}~\bibnamefont {Li}}, \bibinfo {author} {\bibfnamefont
  {X.-H.}\ \bibnamefont {Bao}}, \ and\ \bibinfo {author} {\bibfnamefont
  {J.-W.}\ \bibnamefont {Pan}},\ }\href {\doibase 10.1103/physreva.102.013706}
  {\bibfield  {journal} {\bibinfo  {journal} {Physical Review A}\ }\textbf
  {\bibinfo {volume} {102}},\ \bibinfo {pages} {013706} (\bibinfo {year}
  {2020})}\BibitemShut {NoStop}%
\bibitem [{\citenamefont {Xu.}\ and\ \citenamefont {Coen}(2014)}]{xu}%
  \BibitemOpen
  \bibfield  {author} {\bibinfo {author} {\bibfnamefont {Y.}~\bibnamefont
  {Xu.}}\ and\ \bibinfo {author} {\bibfnamefont {S.}~\bibnamefont {Coen}},\
  }\href@noop {} {\bibfield  {journal} {\bibinfo  {journal} {Opt. Lett.}\
  }\textbf {\bibinfo {volume} {39}},\ \bibinfo {pages} {3492} (\bibinfo {year}
  {2014})}\BibitemShut {NoStop}%
\bibitem [{\citenamefont {Djotyan}\ \emph {et~al.}(2004)\citenamefont
  {Djotyan}, \citenamefont {Bakos}, \citenamefont {Sorlei},\ and\ \citenamefont
  {Szigeti}}]{djotyan}%
  \BibitemOpen
  \bibfield  {author} {\bibinfo {author} {\bibfnamefont {G.~P.}\ \bibnamefont
  {Djotyan}}, \bibinfo {author} {\bibfnamefont {J.~S.}\ \bibnamefont {Bakos}},
  \bibinfo {author} {\bibfnamefont {Z.}~\bibnamefont {Sorlei}}, \ and\ \bibinfo
  {author} {\bibfnamefont {J.}~\bibnamefont {Szigeti}},\ }\href {\doibase
  10.1103/physreva.70.063406} {\bibfield  {journal} {\bibinfo  {journal}
  {Physical Review A}\ }\textbf {\bibinfo {volume} {70}} (\bibinfo {year}
  {2004}),\ 10.1103/physreva.70.063406}\BibitemShut {NoStop}%
\bibitem [{\citenamefont {Simon}\ \emph {et~al.}(2011)\citenamefont {Simon},
  \citenamefont {Belhadj}, \citenamefont {Chatel}, \citenamefont {Amand},
  \citenamefont {Renucci}, \citenamefont {Lemaitre}, \citenamefont {Krebs},
  \citenamefont {Dalgarno}, \citenamefont {Warburton}, \citenamefont {Marie},\
  and\ \citenamefont {et~al.}}]{simon}%
  \BibitemOpen
  \bibfield  {author} {\bibinfo {author} {\bibfnamefont {C.-M.}\ \bibnamefont
  {Simon}}, \bibinfo {author} {\bibfnamefont {T.}~\bibnamefont {Belhadj}},
  \bibinfo {author} {\bibfnamefont {B.}~\bibnamefont {Chatel}}, \bibinfo
  {author} {\bibfnamefont {T.}~\bibnamefont {Amand}}, \bibinfo {author}
  {\bibfnamefont {P.}~\bibnamefont {Renucci}}, \bibinfo {author} {\bibfnamefont
  {A.}~\bibnamefont {Lemaitre}}, \bibinfo {author} {\bibfnamefont
  {O.}~\bibnamefont {Krebs}}, \bibinfo {author} {\bibfnamefont {P.~A.}\
  \bibnamefont {Dalgarno}}, \bibinfo {author} {\bibfnamefont {R.~J.}\
  \bibnamefont {Warburton}}, \bibinfo {author} {\bibfnamefont {X.}~\bibnamefont
  {Marie}}, \ and\ \bibinfo {author} {\bibnamefont {et~al.}},\ }\href {\doibase
  10.1103/physrevlett.106.166801} {\bibfield  {journal} {\bibinfo  {journal}
  {Physical Review Letters}\ }\textbf {\bibinfo {volume} {106}} (\bibinfo
  {year} {2011}),\ 10.1103/physrevlett.106.166801}\BibitemShut {NoStop}%
\bibitem [{\citenamefont {Li}\ \emph {et~al.}(1991)\citenamefont {Li},
  \citenamefont {Bennett},\ and\ \citenamefont {Stedman}}]{li1991}%
  \BibitemOpen
  \bibfield  {author} {\bibinfo {author} {\bibfnamefont {Z.}~\bibnamefont
  {Li}}, \bibinfo {author} {\bibfnamefont {R.}~\bibnamefont {Bennett}}, \ and\
  \bibinfo {author} {\bibfnamefont {G.}~\bibnamefont {Stedman}},\ }\href
  {\doibase 10.1016/0030-4018(91)90242-6} {\bibfield  {journal} {\bibinfo
  {journal} {Optics Communications}\ }\textbf {\bibinfo {volume} {86}},\
  \bibinfo {pages} {51–57} (\bibinfo {year} {1991})}\BibitemShut {NoStop}%
\bibitem [{\citenamefont {Li}\ \emph {et~al.}(1996)\citenamefont {Li},
  \citenamefont {Burkett},\ and\ \citenamefont {Xiao}}]{li1996}%
  \BibitemOpen
  \bibfield  {author} {\bibinfo {author} {\bibfnamefont {Y.-Q.}\ \bibnamefont
  {Li}}, \bibinfo {author} {\bibfnamefont {W.~H.}\ \bibnamefont {Burkett}}, \
  and\ \bibinfo {author} {\bibfnamefont {M.}~\bibnamefont {Xiao}},\ }\href
  {\doibase 10.1364/ol.21.000982} {\bibfield  {journal} {\bibinfo  {journal}
  {Optics Letters}\ }\textbf {\bibinfo {volume} {21}},\ \bibinfo {pages} {982}
  (\bibinfo {year} {1996})}\BibitemShut {NoStop}%
\bibitem [{\citenamefont {Süptitz}\ \emph {et~al.}(1997)\citenamefont
  {Süptitz}, \citenamefont {Duncan},\ and\ \citenamefont {Gould}}]{suptitz}%
  \BibitemOpen
  \bibfield  {author} {\bibinfo {author} {\bibfnamefont {W.}~\bibnamefont
  {Süptitz}}, \bibinfo {author} {\bibfnamefont {B.~C.}\ \bibnamefont
  {Duncan}}, \ and\ \bibinfo {author} {\bibfnamefont {P.~L.}\ \bibnamefont
  {Gould}},\ }\href@noop {} {\bibfield  {journal} {\bibinfo  {journal} {J. Opt.
  Soc. Am. B}\ }\textbf {\bibinfo {volume} {14}},\ \bibinfo {pages} {1001}
  (\bibinfo {year} {1997})}\BibitemShut {NoStop}%
\bibitem [{\citenamefont {Rangelov}\ \emph
  {et~al.}(2005{\natexlab{a}})\citenamefont {Rangelov}, \citenamefont {Piilo},\
  and\ \citenamefont {Vitanov}}]{rangelov2}%
  \BibitemOpen
  \bibfield  {author} {\bibinfo {author} {\bibfnamefont {A.~A.}\ \bibnamefont
  {Rangelov}}, \bibinfo {author} {\bibfnamefont {J.}~\bibnamefont {Piilo}}, \
  and\ \bibinfo {author} {\bibfnamefont {N.~V.}\ \bibnamefont {Vitanov}},\
  }\href {\doibase 10.1103/PhysRevA.72.053404} {\bibfield  {journal} {\bibinfo
  {journal} {Phys. Rev. A}\ }\textbf {\bibinfo {volume} {72}},\ \bibinfo
  {pages} {053404} (\bibinfo {year} {2005}{\natexlab{a}})}\BibitemShut
  {NoStop}%
\bibitem [{\citenamefont {Rangelov}\ \emph
  {et~al.}(2005{\natexlab{b}})\citenamefont {Rangelov}, \citenamefont
  {Vitanov}, \citenamefont {Yatsenko}, \citenamefont {Shore}, \citenamefont
  {Halfmann},\ and\ \citenamefont {Bergmann}}]{rangelov1}%
  \BibitemOpen
  \bibfield  {author} {\bibinfo {author} {\bibfnamefont {A.~A.}\ \bibnamefont
  {Rangelov}}, \bibinfo {author} {\bibfnamefont {N.~V.}\ \bibnamefont
  {Vitanov}}, \bibinfo {author} {\bibfnamefont {L.~P.}\ \bibnamefont
  {Yatsenko}}, \bibinfo {author} {\bibfnamefont {B.~W.}\ \bibnamefont {Shore}},
  \bibinfo {author} {\bibfnamefont {T.}~\bibnamefont {Halfmann}}, \ and\
  \bibinfo {author} {\bibfnamefont {K.}~\bibnamefont {Bergmann}},\ }\href
  {\doibase 10.1103/physreva.72.053403} {\bibfield  {journal} {\bibinfo
  {journal} {Physical Review A}\ }\textbf {\bibinfo {volume} {72}},\ \bibinfo
  {pages} {053403} (\bibinfo {year} {2005}{\natexlab{b}})}\BibitemShut
  {NoStop}%
\bibitem [{\citenamefont {Konar}\ \emph {et~al.}(2012)\citenamefont {Konar},
  \citenamefont {Lozovoy},\ and\ \citenamefont {Dantus}}]{konar}%
  \BibitemOpen
  \bibfield  {author} {\bibinfo {author} {\bibfnamefont {A.}~\bibnamefont
  {Konar}}, \bibinfo {author} {\bibfnamefont {V.~V.}\ \bibnamefont {Lozovoy}},
  \ and\ \bibinfo {author} {\bibfnamefont {M.}~\bibnamefont {Dantus}},\ }\href
  {\doibase 10.1021/jz300761x} {\bibfield  {journal} {\bibinfo  {journal} {The
  Journal of Physical Chemistry Letters}\ }\textbf {\bibinfo {volume} {3}},\
  \bibinfo {pages} {2458} (\bibinfo {year} {2012})},\ \bibinfo {note} {pMID:
  26292133},\ \Eprint {http://arxiv.org/abs/https://doi.org/10.1021/jz300761x}
  {https://doi.org/10.1021/jz300761x} \BibitemShut {NoStop}%
\bibitem [{\citenamefont {Torosov}\ \emph {et~al.}(2010)\citenamefont
  {Torosov}, \citenamefont {Vasilev},\ and\ \citenamefont
  {Vitanov}}]{torosov1}%
  \BibitemOpen
  \bibfield  {author} {\bibinfo {author} {\bibfnamefont {B.}~\bibnamefont
  {Torosov}}, \bibinfo {author} {\bibfnamefont {G.}~\bibnamefont {Vasilev}}, \
  and\ \bibinfo {author} {\bibfnamefont {N.}~\bibnamefont {Vitanov}},\ }\href
  {\doibase 10.1016/j.optcom.2009.11.058} {\bibfield  {journal} {\bibinfo
  {journal} {Optics Communications}\ }\textbf {\bibinfo {volume} {283}},\
  \bibinfo {pages} {1338–1345} (\bibinfo {year} {2010})}\BibitemShut
  {NoStop}%
\bibitem [{\citenamefont {Demkov}\ and\ \citenamefont
  {Kunicke}(1969)}]{demkov}%
  \BibitemOpen
  \bibfield  {author} {\bibinfo {author} {\bibfnamefont {Y.~N.}\ \bibnamefont
  {Demkov}}\ and\ \bibinfo {author} {\bibfnamefont {M.}~\bibnamefont
  {Kunicke}},\ }\href@noop {} {\bibfield  {journal} {\bibinfo  {journal}
  {Vestn. Leningr. Univ. Ser. Fiz. Khim.}\ }\textbf {\bibinfo {volume} {16}},\
  \bibinfo {pages} {39} (\bibinfo {year} {1969})}\BibitemShut {NoStop}%
\bibitem [{\citenamefont {Grimm}\ and\ \citenamefont {Mlynek}(1989)}]{grimm}%
  \BibitemOpen
  \bibfield  {author} {\bibinfo {author} {\bibfnamefont {R.}~\bibnamefont
  {Grimm}}\ and\ \bibinfo {author} {\bibfnamefont {J.}~\bibnamefont {Mlynek}},\
  }\href {\doibase 10.1103/PhysRevLett.63.232} {\bibfield  {journal} {\bibinfo
  {journal} {Phys. Rev. Lett.}\ }\textbf {\bibinfo {volume} {63}},\ \bibinfo
  {pages} {232} (\bibinfo {year} {1989})}\BibitemShut {NoStop}%
\bibitem [{\citenamefont {Oates}\ \emph {et~al.}(2005)\citenamefont {Oates},
  \citenamefont {Wilpers},\ and\ \citenamefont {Hollberg}}]{oates}%
  \BibitemOpen
  \bibfield  {author} {\bibinfo {author} {\bibfnamefont {C.}~\bibnamefont
  {Oates}}, \bibinfo {author} {\bibfnamefont {G.}~\bibnamefont {Wilpers}}, \
  and\ \bibinfo {author} {\bibfnamefont {L.}~\bibnamefont {Hollberg}},\ }\href
  {\doibase 10.1103/physreva.71.023404} {\bibfield  {journal} {\bibinfo
  {journal} {Physical Review A}\ }\textbf {\bibinfo {volume} {71}} (\bibinfo
  {year} {2005}),\ 10.1103/physreva.71.023404}\BibitemShut {NoStop}%
\bibitem [{\citenamefont {Kuznetsova}\ \emph {et~al.}(2014)\citenamefont
  {Kuznetsova}, \citenamefont {Liu},\ and\ \citenamefont
  {Malinovskaya}}]{kuznetsova}%
  \BibitemOpen
  \bibfield  {author} {\bibinfo {author} {\bibfnamefont {E.}~\bibnamefont
  {Kuznetsova}}, \bibinfo {author} {\bibfnamefont {G.}~\bibnamefont {Liu}}, \
  and\ \bibinfo {author} {\bibfnamefont {S.~A.}\ \bibnamefont {Malinovskaya}},\
  }\href {\doibase 10.1088/0031-8949/2014/t160/014024} {\bibfield  {journal}
  {\bibinfo  {journal} {Physica Scripta}\ }\textbf {\bibinfo {volume} {T160}},\
  \bibinfo {pages} {014024} (\bibinfo {year} {2014})}\BibitemShut {NoStop}%
\bibitem [{\citenamefont {Bord\'e}\ \emph {et~al.}(1976)\citenamefont
  {Bord\'e}, \citenamefont {Hall}, \citenamefont {Kunasz},\ and\ \citenamefont
  {Hummer}}]{borde}%
  \BibitemOpen
  \bibfield  {author} {\bibinfo {author} {\bibfnamefont {C.~J.}\ \bibnamefont
  {Bord\'e}}, \bibinfo {author} {\bibfnamefont {J.~L.}\ \bibnamefont {Hall}},
  \bibinfo {author} {\bibfnamefont {C.~V.}\ \bibnamefont {Kunasz}}, \ and\
  \bibinfo {author} {\bibfnamefont {D.~G.}\ \bibnamefont {Hummer}},\ }\href
  {\doibase 10.1103/PhysRevA.14.236} {\bibfield  {journal} {\bibinfo  {journal}
  {Phys. Rev. A}\ }\textbf {\bibinfo {volume} {14}},\ \bibinfo {pages} {236}
  (\bibinfo {year} {1976})}\BibitemShut {NoStop}%
\bibitem [{\citenamefont {Park}\ \emph {et~al.}(2001)\citenamefont {Park},
  \citenamefont {Lee}, \citenamefont {Kwon},\ and\ \citenamefont {Cho}}]{park}%
  \BibitemOpen
  \bibfield  {author} {\bibinfo {author} {\bibfnamefont {S.~E.}\ \bibnamefont
  {Park}}, \bibinfo {author} {\bibfnamefont {H.~S.}\ \bibnamefont {Lee}},
  \bibinfo {author} {\bibfnamefont {T.~Y.}\ \bibnamefont {Kwon}}, \ and\
  \bibinfo {author} {\bibfnamefont {H.}~\bibnamefont {Cho}},\ }\href@noop {}
  {\bibfield  {journal} {\bibinfo  {journal} {Optics Communications}\ }\textbf
  {\bibinfo {volume} {192}},\ \bibinfo {pages} {49} (\bibinfo {year}
  {2001})}\BibitemShut {NoStop}%
\bibitem [{\citenamefont {Hirano}(1988)}]{hirano1}%
  \BibitemOpen
  \bibfield  {author} {\bibinfo {author} {\bibfnamefont {I.}~\bibnamefont
  {Hirano}},\ }\href {\doibase https://doi.org/10.1016/0022-4073(88)90098-2}
  {\bibfield  {journal} {\bibinfo  {journal} {Journal of Quantitative
  Spectroscopy and Radiative Transfer}\ }\textbf {\bibinfo {volume} {39}},\
  \bibinfo {pages} {341} (\bibinfo {year} {1988})}\BibitemShut {NoStop}%
\bibitem [{\citenamefont {LeFloch}\ \emph {et~al.}(1981)\citenamefont
  {LeFloch}, \citenamefont {Lenormand}, \citenamefont {Jezequel},\ and\
  \citenamefont {R.}}]{lefloch}%
  \BibitemOpen
  \bibfield  {author} {\bibinfo {author} {\bibfnamefont {A.}~\bibnamefont
  {LeFloch}}, \bibinfo {author} {\bibfnamefont {J.~M.}\ \bibnamefont
  {Lenormand}}, \bibinfo {author} {\bibfnamefont {G.}~\bibnamefont {Jezequel}},
  \ and\ \bibinfo {author} {\bibfnamefont {L.}~\bibnamefont {R.}},\ }\href@noop
  {} {\bibfield  {journal} {\bibinfo  {journal} {Opt. Lett.}\ }\textbf
  {\bibinfo {volume} {6}},\ \bibinfo {pages} {48} (\bibinfo {year}
  {1981})}\BibitemShut {NoStop}%
\bibitem [{\citenamefont {Sharaby}\ \emph {et~al.}(2013)\citenamefont
  {Sharaby}, \citenamefont {Hassan},\ and\ \citenamefont {Joshi}}]{sharaby}%
  \BibitemOpen
  \bibfield  {author} {\bibinfo {author} {\bibfnamefont {Y.~A.}\ \bibnamefont
  {Sharaby}}, \bibinfo {author} {\bibfnamefont {S.~S.}\ \bibnamefont {Hassan}},
  \ and\ \bibinfo {author} {\bibfnamefont {A.~S.}\ \bibnamefont {Joshi}},\
  }\href@noop {} {\bibfield  {journal} {\bibinfo  {journal} {Journal of
  Nonlinear Physics and Materials}\ }\textbf {\bibinfo {volume} {22}},\
  \bibinfo {pages} {1350044} (\bibinfo {year} {2013})}\BibitemShut {NoStop}%
\bibitem [{\citenamefont {Renzoni}\ \emph {et~al.}(1999)\citenamefont
  {Renzoni}, \citenamefont {Lindner},\ and\ \citenamefont
  {Arimondo}}]{renzoni_lindner_arimondo_1999}%
  \BibitemOpen
  \bibfield  {author} {\bibinfo {author} {\bibfnamefont {F.}~\bibnamefont
  {Renzoni}}, \bibinfo {author} {\bibfnamefont {A.}~\bibnamefont {Lindner}}, \
  and\ \bibinfo {author} {\bibfnamefont {E.}~\bibnamefont {Arimondo}},\ }\href
  {\doibase 10.1103/physreva.60.450} {\bibfield  {journal} {\bibinfo  {journal}
  {Physical Review A}\ }\textbf {\bibinfo {volume} {60}},\ \bibinfo {pages}
  {450–455} (\bibinfo {year} {1999})}\BibitemShut {NoStop}%
\bibitem [{\citenamefont {Valente}\ \emph {et~al.}(2003)\citenamefont
  {Valente}, \citenamefont {Failache},\ and\ \citenamefont {Lezama}}]{valente}%
  \BibitemOpen
  \bibfield  {author} {\bibinfo {author} {\bibfnamefont {P.}~\bibnamefont
  {Valente}}, \bibinfo {author} {\bibfnamefont {H.}~\bibnamefont {Failache}}, \
  and\ \bibinfo {author} {\bibfnamefont {A.}~\bibnamefont {Lezama}},\ }\href
  {\doibase 10.1103/PhysRevA.67.013806} {\bibfield  {journal} {\bibinfo
  {journal} {Phys. Rev. A}\ }\textbf {\bibinfo {volume} {67}},\ \bibinfo
  {pages} {013806} (\bibinfo {year} {2003})}\BibitemShut {NoStop}%
\bibitem [{\citenamefont {Phillips}(2021)}]{phillips2}%
  \BibitemOpen
  \bibfield  {author} {\bibinfo {author} {\bibfnamefont {D.~F.}\ \bibnamefont
  {Phillips}},\ }\href@noop {} {\  (\bibinfo {year} {2021})}\BibitemShut
  {NoStop}%
\bibitem [{\citenamefont {Phillips}\ \emph {et~al.}(2005)\citenamefont
  {Phillips}, \citenamefont {Novikova}, \citenamefont {Wang}, \citenamefont
  {Walsworth},\ and\ \citenamefont {Crescimanno}}]{phillips}%
  \BibitemOpen
  \bibfield  {author} {\bibinfo {author} {\bibfnamefont {D.~F.}\ \bibnamefont
  {Phillips}}, \bibinfo {author} {\bibfnamefont {I.}~\bibnamefont {Novikova}},
  \bibinfo {author} {\bibfnamefont {C.~Y.-T.}\ \bibnamefont {Wang}}, \bibinfo
  {author} {\bibfnamefont {R.~L.}\ \bibnamefont {Walsworth}}, \ and\ \bibinfo
  {author} {\bibfnamefont {M.}~\bibnamefont {Crescimanno}},\ }\href@noop {}
  {\bibfield  {journal} {\bibinfo  {journal} {J. Opt. Soc. Am. B}\ }\textbf
  {\bibinfo {volume} {22}},\ \bibinfo {pages} {305} (\bibinfo {year}
  {2005})}\BibitemShut {NoStop}%
\bibitem [{\citenamefont {Rubbmark}\ \emph {et~al.}(1981)\citenamefont
  {Rubbmark}, \citenamefont {Kash}, \citenamefont {Littman},\ and\
  \citenamefont {Kleppner}}]{rubbmark_kash_littman_kleppner_1981}%
  \BibitemOpen
  \bibfield  {author} {\bibinfo {author} {\bibfnamefont {J.~R.}\ \bibnamefont
  {Rubbmark}}, \bibinfo {author} {\bibfnamefont {M.~M.}\ \bibnamefont {Kash}},
  \bibinfo {author} {\bibfnamefont {M.~G.}\ \bibnamefont {Littman}}, \ and\
  \bibinfo {author} {\bibfnamefont {D.}~\bibnamefont {Kleppner}},\ }\href
  {\doibase 10.1103/physreva.23.3107} {\bibfield  {journal} {\bibinfo
  {journal} {Physical Review A}\ }\textbf {\bibinfo {volume} {23}},\ \bibinfo
  {pages} {3107–3117} (\bibinfo {year} {1981})}\BibitemShut {NoStop}%
\bibitem [{\citenamefont {Shytov}(2004)}]{Shytov}%
  \BibitemOpen
  \bibfield  {author} {\bibinfo {author} {\bibfnamefont {A.~V.}\ \bibnamefont
  {Shytov}},\ }\href {\doibase 10.1103/PhysRevA.70.052708} {\bibfield
  {journal} {\bibinfo  {journal} {Phys. Rev. A}\ }\textbf {\bibinfo {volume}
  {70}},\ \bibinfo {pages} {052708} (\bibinfo {year} {2004})}\BibitemShut
  {NoStop}%
\bibitem [{\citenamefont {Kitamura}\ \emph {et~al.}(2020)\citenamefont
  {Kitamura}, \citenamefont {Nagoasa},\ and\ \citenamefont
  {Morimoto}}]{kitamura}%
  \BibitemOpen
  \bibfield  {author} {\bibinfo {author} {\bibfnamefont {S.}~\bibnamefont
  {Kitamura}}, \bibinfo {author} {\bibfnamefont {N.}~\bibnamefont {Nagoasa}}, \
  and\ \bibinfo {author} {\bibfnamefont {T.}~\bibnamefont {Morimoto}},\
  }\href@noop {} {\bibfield  {journal} {\bibinfo  {journal} {Communications
  Physics}\ }\textbf {\bibinfo {volume} {3}},\ \bibinfo {pages} {2} (\bibinfo
  {year} {2020})}\BibitemShut {NoStop}%
\bibitem [{\citenamefont {Akulin}\ and\ \citenamefont
  {Schleich}(1992)}]{Akulin}%
  \BibitemOpen
  \bibfield  {author} {\bibinfo {author} {\bibfnamefont {V.~M.}\ \bibnamefont
  {Akulin}}\ and\ \bibinfo {author} {\bibfnamefont {W.~P.}\ \bibnamefont
  {Schleich}},\ }\href {\doibase 10.1103/PhysRevA.46.4110} {\bibfield
  {journal} {\bibinfo  {journal} {Phys. Rev. A}\ }\textbf {\bibinfo {volume}
  {46}},\ \bibinfo {pages} {4110} (\bibinfo {year} {1992})}\BibitemShut
  {NoStop}%
\bibitem [{\citenamefont {Avishai}\ and\ \citenamefont {Band}(2014)}]{Avishai}%
  \BibitemOpen
  \bibfield  {author} {\bibinfo {author} {\bibfnamefont {Y.}~\bibnamefont
  {Avishai}}\ and\ \bibinfo {author} {\bibfnamefont {Y.~B.}\ \bibnamefont
  {Band}},\ }\href {\doibase 10.1103/physreva.90.032116} {\bibfield  {journal}
  {\bibinfo  {journal} {Physical Review A}\ }\textbf {\bibinfo {volume} {90}}
  (\bibinfo {year} {2014}),\ 10.1103/physreva.90.032116}\BibitemShut {NoStop}%
\bibitem [{\citenamefont {Shore}\ \emph {et~al.}(2009)\citenamefont {Shore},
  \citenamefont {Gromovyy}, \citenamefont {Yatsenko},\ and\ \citenamefont
  {Romanenko}}]{shore_gromovyy_yatsenko_romanenko_2009}%
  \BibitemOpen
  \bibfield  {author} {\bibinfo {author} {\bibfnamefont {B.~W.}\ \bibnamefont
  {Shore}}, \bibinfo {author} {\bibfnamefont {M.~V.}\ \bibnamefont {Gromovyy}},
  \bibinfo {author} {\bibfnamefont {L.~P.}\ \bibnamefont {Yatsenko}}, \ and\
  \bibinfo {author} {\bibfnamefont {V.~I.}\ \bibnamefont {Romanenko}},\ }\href
  {\doibase 10.1119/1.3231688} {\bibfield  {journal} {\bibinfo  {journal}
  {American Journal of Physics}\ }\textbf {\bibinfo {volume} {77}},\ \bibinfo
  {pages} {1183–1194} (\bibinfo {year} {2009})}\BibitemShut {NoStop}%
\bibitem [{\citenamefont {Ernst}(1966)}]{ernst}%
  \BibitemOpen
  \bibfield  {author} {\bibinfo {author} {\bibfnamefont {R.~R.}\ \bibnamefont
  {Ernst}},\ }\href@noop {} {\bibfield  {journal} {\bibinfo  {journal}
  {Advances in Magnetic Resonance}\ }\textbf {\bibinfo {volume} {2}},\ \bibinfo
  {pages} {1–135} (\bibinfo {year} {1966})}\BibitemShut {NoStop}%
\bibitem [{\citenamefont {White}\ \emph {et~al.}(2021)\citenamefont {White},
  \citenamefont {Pearce}, \citenamefont {Vagie},\ and\ \citenamefont
  {Kusenko}}]{white}%
  \BibitemOpen
  \bibfield  {author} {\bibinfo {author} {\bibfnamefont {G.}~\bibnamefont
  {White}}, \bibinfo {author} {\bibfnamefont {L.}~\bibnamefont {Pearce}},
  \bibinfo {author} {\bibfnamefont {D.}~\bibnamefont {Vagie}}, \ and\ \bibinfo
  {author} {\bibfnamefont {A.}~\bibnamefont {Kusenko}},\ }\href {\doibase
  10.1103/PhysRevLett.127.181601} {\bibfield  {journal} {\bibinfo  {journal}
  {Phys. Rev. Lett.}\ }\textbf {\bibinfo {volume} {127}},\ \bibinfo {pages}
  {181601} (\bibinfo {year} {2021})}\BibitemShut {NoStop}%
\bibitem [{\citenamefont {Torosov}\ and\ \citenamefont
  {Vitanov}(2011)}]{torosov2}%
  \BibitemOpen
  \bibfield  {author} {\bibinfo {author} {\bibfnamefont {B.~T.}\ \bibnamefont
  {Torosov}}\ and\ \bibinfo {author} {\bibfnamefont {N.~V.}\ \bibnamefont
  {Vitanov}},\ }\href@noop {} {\bibfield  {journal} {\bibinfo  {journal} {Phys.
  Rev. A}\ }\textbf {\bibinfo {volume} {84}},\ \bibinfo {pages} {063411}
  (\bibinfo {year} {2011})}\BibitemShut {NoStop}%
\bibitem [{SAS()}]{SASVapor}%
  \BibitemOpen
  \href@noop {} {\bibinfo  {journal} {SAS is typically done at room
  temperature. We worked at this slightly elevated temperature to have better
  signal to noise and to lock the vapor density thus avoiding opacity drifts
  that could complicate the identification of chirp asymmetry in the probe
  transmission signal.}\ }\BibitemShut {NoStop}%
\bibitem [{\citenamefont {Wizemann}\ and\ \citenamefont
  {Niemax}(1998)}]{wizemann}%
  \BibitemOpen
\bibfield  {journal} {  }\bibfield  {author} {\bibinfo {author} {\bibfnamefont
  {H.~D.}\ \bibnamefont {Wizemann}}\ and\ \bibinfo {author} {\bibfnamefont
  {K.}~\bibnamefont {Niemax}},\ }\href@noop {} {\bibfield  {journal} {\bibinfo
  {journal} {Microchimica Acta}\ }\textbf {\bibinfo {volume} {129}},\ \bibinfo
  {pages} {209} (\bibinfo {year} {1998})}\BibitemShut {NoStop}%
\bibitem [{\citenamefont {Bucci}\ \emph {et~al.}(2021)\citenamefont {Bucci},
  \citenamefont {Feigert}, \citenamefont {Crescimanno}, \citenamefont
  {Chamberlain},\ and\ \citenamefont {Giovannone}}]{bucci}%
  \BibitemOpen
  \bibfield  {author} {\bibinfo {author} {\bibfnamefont {T.~J.}\ \bibnamefont
  {Bucci}}, \bibinfo {author} {\bibfnamefont {J.}~\bibnamefont {Feigert}},
  \bibinfo {author} {\bibfnamefont {M.}~\bibnamefont {Crescimanno}}, \bibinfo
  {author} {\bibfnamefont {B.}~\bibnamefont {Chamberlain}}, \ and\ \bibinfo
  {author} {\bibfnamefont {A.}~\bibnamefont {Giovannone}},\ }\href@noop {}
  {\bibfield  {journal} {\bibinfo  {journal} {Am. Journal of Phys.}\ }\textbf
  {\bibinfo {volume} {89}},\ \bibinfo {pages} {730} (\bibinfo {year}
  {2021})}\BibitemShut {NoStop}%
\bibitem [{\citenamefont {Frank}(1953)}]{frank}%
  \BibitemOpen
  \bibfield  {author} {\bibinfo {author} {\bibfnamefont {F.~C.}\ \bibnamefont
  {Frank}},\ }\href@noop {} {\bibfield  {journal} {\bibinfo  {journal}
  {Biochimica et Biophysica Acta}\ }\textbf {\bibinfo {volume} {11}},\ \bibinfo
  {pages} {459} (\bibinfo {year} {1953})}\BibitemShut {NoStop}%
\bibitem [{\citenamefont {Breveglieri}\ \emph {et~al.}(2018)\citenamefont
  {Breveglieri}, \citenamefont {Maggioni},\ and\ \citenamefont
  {Mazzotti}}]{Breveglieri}%
  \BibitemOpen
  \bibfield  {author} {\bibinfo {author} {\bibfnamefont {F.}~\bibnamefont
  {Breveglieri}}, \bibinfo {author} {\bibfnamefont {G.~M.}\ \bibnamefont
  {Maggioni}}, \ and\ \bibinfo {author} {\bibfnamefont {M.}~\bibnamefont
  {Mazzotti}},\ }\href@noop {} {\bibfield  {journal} {\bibinfo  {journal}
  {Cryst. Growth Des.}\ }\textbf {\bibinfo {volume} {18}},\ \bibinfo {pages}
  {1873} (\bibinfo {year} {2018})}\BibitemShut {NoStop}%
\end{thebibliography}%

\end{document}